\documentclass[10pt]{article}
\usepackage{graphicx}
\usepackage{amsmath}
\usepackage{amssymb}
\usepackage{caption2}
\begin{document}
\bibliographystyle{prsty}
\begin{center}
{\large {\bf \sc{  Analysis of  the $QQ\bar{Q}\bar{Q}$  tetraquark states with  QCD sum rules }}} \\[2mm]
Zhi-Gang  Wang \footnote{E-mail: zgwang@aliyun.com.  }   \\
 Department of Physics, North China Electric Power University, Baoding 071003, P. R. China
\end{center}

\begin{abstract}
In this article, we study the $J^{PC}=0^{++}$ and $2^{++}$ $QQ\bar{Q}\bar{Q}$ tetraquark states with the QCD sum rules, and  obtain the predictions
$M_{X(cc\bar{c}\bar{c},0^{++})} =5.99\pm0.08\,\rm{GeV}$, $M_{X(cc\bar{c}\bar{c},2^{++})} =6.09\pm0.08\,\rm{GeV}$,
$M_{X(bb\bar{b}\bar{b},0^{++})} =18.84\pm0.09\,\rm{GeV}$ and $M_{X(bb\bar{b}\bar{b},2^{++})}  =18.85\pm0.09\,\rm{GeV}$, which can be confronted to the experimental data in the future. Furthermore, we illustrate that the
diquark-antidiquark type tetraquark state can be  taken as a special superposition of a series of  meson-meson pairs and embodies  the net effects.
\end{abstract}



\section{Introduction}

The observations of the charmonium-like and bottomonium-like  states  have  provided us with a good  opportunity   to study the
  exotic states and understand the strong interactions, especially those charged  states    $Z_c(3885)$, $Z_c(3900)$,  $Z_c(4020)$, $Z_c(4025)$, $Z_c(4200)$, $Z(4430)$,
$Z_b(10610)$,  $Z_b(10650)$, they are excellent candidates   for  the multiquark states \cite{Zhu-PRT}.
If they are tetraquark states, they consist two heavy quarks and two light quarks, we have to deal with both the heavy and light degrees of freedom of the dynamics.
On the other hand, if there exist tetraquark  configurations consist of four heavy quarks, the dynamics is much simple. There have been several works on the mass spectrum of the $QQ\bar{Q}\bar{Q}$ tetraquark states, such as the non-relativistic potential models \cite{Silvestre-1986,Lloyd-2004,Barnea-2006,Bai-2016},  the  Bethe-Salpeter equation \cite{Heupel-2012}, the constituent diquark model with spin-spin interaction \cite{Berezhnoy-2012,Rosner-2016}, the constituent quark model with color-magnetic interaction \cite{Wu-2016}, the moment QCD sum rules \cite{Chen-2016}, etc. In this article, we study the tetraquark  states consist of  four heavy quarks with the Borel QCD sum rules.

 The  QCD sum rules is a powerful  theoretical tool in studying the ground state tetraquark states and molecular states, and has given many successful descriptions of the masses and hadronic coupling constants \cite{QCDSR-Tetraquark-Molecule}. In this article, we study the $J^{PC}=0^{++}$ and $2^{++}$ $QQ\bar{Q}\bar{Q}$ tetraquark states, which may be observed in the
 $e^+e^-$ and $pp$ collisions, for example, $e^+e^- \to J/\psi \bar{c}c$, $pp\to \bar{c}c\bar{c}c$.
The ATLAS, CMS and LHCb collaborations have measured the cross section for double charmonium
production \cite{Exp-psipsi}, the CMS collaboration  has observed the $\Upsilon$ pair production \cite{Exp-UpsilonUpsilon}.

The quarks have color $SU(3)$ symmetry,  we can construct the tetraquark states according to the routine  ${\rm quark}\to {\rm diquark}\to {\rm tetraquark}$,
\begin{eqnarray}
({\bf 3}_c\otimes {\bf 3}_c)\otimes(\overline{{\bf 3}}_c\otimes \overline{{\bf 3}}_c) &\to&(\overline{{\bf 3}}_c\oplus {\bf 6}_c)\otimes({\bf 3}_c\oplus \overline{{\bf 6}}_c)\to(\overline{{\bf 3}}_c\otimes {\bf 3}_c)\oplus ({\bf 6}_c\otimes\overline{{\bf 6}}_c) \to({\bf 1}_c\oplus{\bf 8}_c) \oplus \cdots \, ,
\end{eqnarray}
where the ${\bf 1}_c$, ${\bf 3}_c$ ($\overline{{\bf 3}}_c$), ${\bf 6}_c$  and ${\bf 8}_c$ denote the color singlet, triplet (antitriplet), sextet  and octet, respectively.
The  one-gluon exchange leads to attractive (repulsive) interaction in the color   antitriplet (sextet) channel, which favors (disfavors) the formation of diquark states in the color antitriplet (sextet) \cite{One-gluon}.
The diquarks $\varepsilon^{ijk} q^{T}_j C\Gamma q^{\prime}_k$ in color
 antitriplet have  five  structures  in Dirac spinor space, where the $i$, $j$ and $k$ are color indexes, $C\Gamma=C\gamma_5$, $C$, $C\gamma_\mu \gamma_5$,  $C\gamma_\mu $ and $C\sigma_{\mu\nu}$ for the scalar, pseudoscalar, vector, axialvector  and  tensor diquarks, respectively.
The  stable diquark configurations are the scalar ($C\gamma_5$) and axialvector ($C\gamma_\mu$) diquark states from the QCD sum rules \cite{WangDiquark,WangLDiquark}. The double-heavy diquark states $\varepsilon^{ijk} Q^{T}_j C\gamma_5 Q_k$ cannot exist due to the Pauli principle.
 In this article, we take the double-heavy diquark states $\varepsilon^{ijk} Q^{T}_j C\gamma_\mu Q_k$ as basic constituents to construct the tetraquark states.

The article is arranged as follows:  we derive the QCD sum rules for the masses and pole residues of  the   $QQ\bar{Q}\bar{Q}$ tetraquark states in section 2; in section 3, we present the numerical results and discussions; section 4 is reserved for our conclusion.

\section{QCD sum rules for  the   $QQ\bar{Q}\bar{Q}$ tetraquark states  }
We write down  the two-point correlation functions  $\Pi_{\mu\nu\alpha\beta}(p)$ and $\Pi (p)$ in the QCD sum rules,
\begin{eqnarray}
\Pi_{\mu\nu\alpha\beta}(p)&=&i\int d^4x e^{ip \cdot x} \langle0|T\left\{J_{\mu\nu}(x)J_{\alpha\beta}^{\dagger}(0)\right\}|0\rangle \, , \nonumber\\
\Pi(p)&=&i\int d^4x e^{ip \cdot x} \langle0|T\left\{J(x)J^{\dagger}(0)\right\}|0\rangle \, ,
\end{eqnarray}
where
\begin{eqnarray}
J_{\mu\nu}(x)&=&\frac{\varepsilon^{ijk}\varepsilon^{imn}}{\sqrt{2}}\left\{Q^j(x)C\gamma_\mu Q^k(x) \bar{Q}^m(x) \gamma_\nu C \bar{Q}^n(x)+ Q^j(x)C\gamma_\nu  Q^k(x) \bar{Q}^m(x)\gamma_\mu C \bar{Q}^n(x) \right\} \, , \nonumber \\
J(x)&=&\varepsilon^{ijk}\varepsilon^{imn}Q^j(x)C\gamma_\mu Q^k(x) \bar{Q}^m(x)\gamma^\mu C \bar{Q}^n(x) \, ,
\end{eqnarray}
 $Q=c,b$, the $i$, $j$, $k$, $m$, $n$ are color indexes, the $C$ is the charge conjunction matrix. We choose  the   currents $J(x)$ and $J_{\mu\nu}(x)$ to interpolate the $J^{PC}=0^{++}$ and $2^{++}$ diquark-antidiquark type $QQ\bar{Q}\bar{Q}$ tetraquark states, respectively.
In Ref.\cite{Chen-2016}, Chen et al choose the currents $\eta^i(x)$ and $\eta^j_{\mu\nu}(x)$ with $i=1,2,3,4,5$ and $j=1,2$ to interpolate the $0^{++}$ and $2^{++}$ $QQ\bar{Q}\bar{Q}$ tetraquark states, respectively,
\begin{eqnarray}
\eta^i(x)&=&Q^a(x)C\Gamma^i Q^b(x) \bar{Q}^a(x)\Gamma^i C \bar{Q}^b(x)\, , \nonumber\\
\eta^j_{\mu\nu}(x)&=&Q^a(x)C\Gamma^j_\mu Q^b(x) \bar{Q}^a(x)\Gamma^j_\nu C \bar{Q}^b(x)+Q^a(x)C\Gamma^j_\nu Q^b(x) \bar{Q}^a(x)\Gamma^j_\mu C \bar{Q}^b(x)\, ,
\end{eqnarray}
where
$\Gamma^1=\gamma_5$, $\Gamma^2=\gamma_\mu\gamma_5$, $\Gamma^3=\sigma_{\mu\nu}$, $\Gamma^4=\gamma_\mu$, $\Gamma^5=1$, $\Gamma^1_\mu=\gamma_\mu$, $\Gamma^2_\mu=\gamma_\mu\gamma_5$, the $a$ and $b$ are color indexes. The $C\gamma_5$, $C$, $C\gamma_\mu\gamma_5$ are antisymmetric, while the $C\gamma_\mu$, $C\sigma_{\mu\nu}$ are symmetric. So the currents $\eta^{1/2/5}(x)$ and $\eta^2_{\mu\nu}(x)$ are in color ${\bf 6}_c\otimes {\bf \bar 6}_c$ representation, while the currents $\eta^{3/4}(x)$ and $\eta^1_{\mu\nu}(x)$ have both color ${\bf 6}_c\otimes {\bf \bar 6}_c$ and ${\bf \bar{3}}_c\otimes {\bf 3}_c$ components.
The  currents $J(x)$ and $J_{\mu\nu}(x)$  chosen in this article are in the color ${\bf \bar{3}}_c\otimes {\bf 3}_c$ representation, and significantly differ from the currents $\eta^{4}(x)$ and $\eta^1_{\mu\nu}(x)$ chosen in Ref.\cite{Chen-2016}, respectively. The one-gluon exchange leads to attractive (repulsive) interaction in the color
${\bf\bar  3}_c$ ($ { \bf 6}_c$) channel \cite{One-gluon},  the currents or quark structures chosen in the present work and in Ref.\cite{Chen-2016} couple potentially to the tetraquark states with different masses.

At the phenomenological side, we can insert  a complete set of intermediate hadronic states with
the same quantum numbers as the current operators $J_{\mu\nu}(x)$ and $J(x)$ into the
correlation functions $\Pi_{\mu\nu\alpha\beta}(p)$ and $\Pi(p)$ to obtain the hadronic representation
\cite{SVZ79,Reinders85}. After isolating the ground state
contributions of the scalar  and tensor $QQ\bar{Q}\bar{Q}$  tetraquark states (denoted by $X$), we get the following results,
\begin{eqnarray}
\Pi_{\mu\nu\alpha\beta} (p) &=&\frac{\lambda_X^2}{M_X^2-p^2}\left( \frac{\widetilde{g}_{\mu\alpha}\widetilde{g}_{\nu\beta}+\widetilde{g}_{\mu\beta}\widetilde{g}_{\nu\alpha}}{2}-\frac{\widetilde{g}_{\mu\nu}\widetilde{g}_{\alpha\beta}}{3}\right) +\cdots \, \, , \\
\Pi (p) &=&\frac{\lambda_X^2}{M^2_X-p^2} +\cdots \, \, ,
\end{eqnarray}
where $\widetilde{g}_{\mu\nu}=g_{\mu\nu}-\frac{p_{\mu}p_{\nu}}{p^2}$, the pole residues  $\lambda_{X}$ are defined by
\begin{eqnarray}
 \langle 0|J_{\mu\nu}(0)|X (p)\rangle &=& \lambda_{X} \, \varepsilon_{\mu\nu}(\lambda,p)   \, , \nonumber\\
 \langle 0|J (0)|X (p)\rangle &=& \lambda_{X}     \, ,
\end{eqnarray}
the $\varepsilon_{\mu\nu}(\lambda,p)$ is the  polarization vector of the tensor tetraquark states,
 \begin{eqnarray}
 \sum_{\lambda}\varepsilon^*_{\alpha\beta}(\lambda,p)\varepsilon_{\mu\nu}(\lambda,p)
 &=&\frac{\widetilde{g}_{\alpha\mu}\widetilde{g}_{\beta\nu}+\widetilde{g}_{\alpha\nu}\widetilde{g}_{\beta\mu}}{2}-\frac{\widetilde{g}_{\alpha\beta}\widetilde{g}_{\mu\nu}}{3}\,.
 \end{eqnarray}

 Now  we briefly outline  the operator product expansion for the correlation functions $\Pi_{\mu\nu\alpha\beta}(p)$ and $\Pi(p)$ in perturbative QCD.  We contract the heavy quark fields in the correlation functions
$\Pi_{\mu\nu\alpha\beta}(p)$ and $\Pi(p)$ with Wick theorem, and obtain the results:
\begin{eqnarray}
\Pi_{\mu\nu\alpha\beta}(p)&=&\frac{i\varepsilon^{ijk}\varepsilon^{imn}\varepsilon^{i^{\prime}j^{\prime}k^{\prime}}\varepsilon^{i^{\prime}m^{\prime}n^{\prime}}}{2}\int d^4x e^{ip \cdot x}   \nonumber\\
&&\left\{{\rm Tr}\left[ \gamma_{\mu}S^{kk^{\prime}}(x)\gamma_{\alpha} CS^{jj^{\prime}T}(x)C\right] {\rm Tr}\left[ \gamma_{\beta} S^{n^{\prime}n}(-x)\gamma_{\nu} C S^{m^{\prime}mT}(-x)C\right] \right. \nonumber\\
&&+{\rm Tr}\left[ \gamma_{\nu} S^{kk^{\prime}}(x)\gamma_{\beta} CS^{jj^{\prime}T}(x)C\right] {\rm Tr}\left[ \gamma_{\alpha} S^{n^{\prime}n}(-x)\gamma_{\mu} C S^{m^{\prime}mT}(-x)C\right] \nonumber\\
&&+{\rm Tr}\left[ \gamma_{\mu} S^{kk^{\prime}}(x) \gamma_{\beta} CS^{jj^{\prime}T}(x)C\right] {\rm Tr}\left[ \gamma_{\alpha} S^{n^{\prime}n}(-x) \gamma_{\nu}C S^{m^{\prime}mT}(-x)C\right] \nonumber\\
 &&\left.+{\rm Tr}\left[ \gamma_{\nu} S^{kk^{\prime}}(x)\gamma_{\alpha} CS^{jj^{\prime}T}(x)C\right] {\rm Tr}\left[ \gamma_{\beta} S^{n^{\prime}n}(-x)\gamma_{\mu} C S^{m^{\prime}mT}(-x)C\right] \right\} \, , \nonumber\\
 \Pi(p)&=&i\varepsilon^{ijk}\varepsilon^{imn}\varepsilon^{i^{\prime}j^{\prime}k^{\prime}}\varepsilon^{i^{\prime}m^{\prime}n^{\prime}}\int d^4x e^{ip \cdot x}   \nonumber\\
&&{\rm Tr}\left[ \gamma_{\mu}S^{kk^{\prime}}(x)\gamma_{\alpha} CS^{jj^{\prime}T}(x)C\right] {\rm Tr}\left[ \gamma^{\alpha} S^{n^{\prime}n}(-x)\gamma^{\mu} C S^{m^{\prime}mT}(-x)C\right]   \, ,
\end{eqnarray}
where the   $S_{ij}(x)$   is the full $Q$ quark propagator,
 \begin{eqnarray}
S_{ij}(x)&=&\frac{i}{(2\pi)^4}\int d^4k e^{-ik \cdot x} \left\{
\frac{\delta_{ij}}{\!\not\!{k}-m_Q}
-\frac{g_sG^n_{\alpha\beta}t^n_{ij}}{4}\frac{\sigma^{\alpha\beta}(\!\not\!{k}+m_Q)+(\!\not\!{k}+m_Q)
\sigma^{\alpha\beta}}{(k^2-m_Q^2)^2}\right.\nonumber\\
&&\left.+\frac{g_s^2G^n_{\alpha\beta}G^{n\alpha\beta}}{12} \delta_{ij}m_Q \frac{k^2+m_Q\!\not\!{k}}{(k^2-m_Q^2)^4}
+\cdots\right\} \, ,
\end{eqnarray}
and  $t^n=\frac{\lambda^n}{2}$, the $\lambda^n$ is the Gell-Mann matrix \cite{Reinders85}. Then we compute  the integrals both in the coordinate and momentum spaces to obtain the correlation functions $\Pi_{\mu\nu\alpha\beta}(p)$ and $\Pi(p)$, therefore the QCD spectral densities through dispersion relation. The calculations are straightforward but tedious.

 We  take the quark-hadron duality below the continuum thresholds  $s_0$ and perform Borel transform  with respect to
the variable $P^2=-p^2$ to obtain  the QCD sum rules:
\begin{eqnarray}
\lambda^2_{X}\, \exp\left(-\frac{M^2_{X}}{T^2}\right)= \int_{16m_Q^2}^{s_0} ds \int_{z_i}^{z_f}dz \int_{t_i}^{t_f}dt \int_{r_i}^{r_f}dr\, \rho(s,z,t,r)  \, \exp\left(-\frac{s}{T^2}\right) \, ,
\end{eqnarray}
where $\rho(s,z,t,r) =\rho_S(s,z,t,r) $ and $\rho_T(s,z,t,r) $ for the scalar and tensor tetraquark states, respectively, the explicit expressions are given in the Appendix,
\begin{eqnarray}
r_{f/i}&=&\frac{1}{2}\left\{1-z-t \pm \sqrt{(1-z-t)^2-4\frac{1-z-t}{\hat{s}-\frac{1}{z}-\frac{1}{t}}}\right\} \, ,\nonumber\\
t_{f/i}&=&\frac{1}{2\left( \hat{s}-\frac{1}{z}\right)}\left\{ (1-z)\left( \hat{s}-\frac{1}{z}\right)-3 \pm \sqrt{ \left[  (1-z)\left( \hat{s}-\frac{1}{z}\right)-3\right]^2-4 (1-z)\left( \hat{s}-\frac{1}{z}\right)  }\right\}\, ,\nonumber\\
z_{f/i}&=&\frac{1}{2\hat{s}}\left\{ \hat{s}-8 \pm \sqrt{\left(\hat{s}-8\right)^2-4\hat{s}  }\right\}\, ,
\end{eqnarray}
and $\hat{s}=\frac{s}{m_Q^2}$.

 We derive   Eq.(11) with respect to  $\tau=\frac{1}{T^2}$, then eliminate the
 pole residues $\lambda_{X}$, and  obtain the QCD sum rules for
 the masses of the scalar   and tensor $QQ\bar{Q}\bar{Q}$ tetraquark states,
 \begin{eqnarray}
 M^2_{X}&=&- \frac{\frac{d}{d \tau}\int_{16m_Q^2}^{s_0} ds \int_{z_i}^{z_f}dz \int_{t_i}^{t_f}dt \int_{r_i}^{r_f}dr\, \rho(s,z,t,r)  \, \exp\left(-\tau s\right)}{\int_{16m_Q^2}^{s_0} ds \int_{z_i}^{z_f}dz \int_{t_i}^{t_f}dt \int_{r_i}^{r_f}dr\, \rho(s,z,t,r)  \, \exp\left(-\tau s\right)}\, .
\end{eqnarray}

In the moment QCD sum rules, the moments $\overline{M}_n(P_0^2)$ at the phenomenological side are defined by
\begin{eqnarray}
\overline{M}_n(P_0^2)&=&\frac{\lambda_X^2}{(M_X^2+P_0^2)^{n+1}}+\frac{\lambda_{X^\prime}^2}{(M_X^{\prime2}+P_0^2)^{n+1}}+\cdots \, , \nonumber\\
&=&\frac{\lambda_X^2}{(M_X^2+P_0^2)^{n+1}}\left[1+ \frac{\lambda_{X^\prime}^2}{\lambda_X^2}\frac{(M_X^2+P_0^2)^{n+1}}{(M_X^{\prime2}+P_0^2)^{n+1}}+\cdots\right]\, , \nonumber\\
&=&\frac{\lambda_X^2}{(M_X^2+P_0^2)^{n+1}}\left[1+\delta_n(P_0^2)\right]\, ,
\end{eqnarray}
where the $X^\prime$ denotes the first radial excited state of the $X$, the $P_0^2$ is a particular value for the parameter $P^2=-p^2$.
We can extract the mass $M_X$ according to the ratio  $r(n,P_0^2)$ at large values of $n$,
\begin{eqnarray}
r(n,P_0^2)&=&\frac{\overline{M}_n(P_0^2)}{\overline{M}_{n+1}(P_0^2)}=(M_X^2+P_0^2)\frac{1+\delta_n(P_0^2)}{1+\delta_{n+1}(P_0^2)}\, ,
\end{eqnarray}
where the small values $\delta_{n}(P_0^2)\approx \delta_{n+1}(P_0^2)$. In Refs.\cite{Wang4430,Wang4500}, we observe that $\frac{\lambda_{X^\prime}^2}{\lambda_X^2}=6.6$ (or $9.6$) for the central values of the pole residues for $X=Z_c(3900)$, $X^\prime=Z(4430)$ (or $X=X(3915)$, $X^\prime=X(4500)$) in the  scenario of tetraquark  states. So the $n$ has to be postponed to very large values \cite{Chen-2016}. In the present work, the contributions of the high resonances and continuum states are depressed by the weight function $\exp\left( -\frac{s}{T^2}\right)$. The differences between  the predicted masses in the present work and in Ref.\cite{Chen-2016} originate from the different currents or quark structures.

\section{Numerical results and discussions}

We take the gluon condensate  to be the standard value
\cite{SVZ79,Reinders85,ColangeloReview}, and  take the $\overline{MS}$ masses $m_{c}(m_c)=(1.275\pm0.025)\,\rm{GeV}$ and $m_{b}(m_b)=(4.18\pm0.03)\,\rm{GeV}$
 from the Particle Data Group \cite{PDG}.
We take into account
the energy-scale dependence of  the  $\overline{MS}$ masses from the renormalization group equation,
 \begin{eqnarray}
m_c(\mu)&=&m_c(m_c)\left[\frac{\alpha_{s}(\mu)}{\alpha_{s}(m_c)}\right]^{\frac{12}{25}} \, ,\nonumber\\
m_b(\mu)&=&m_b(m_b)\left[\frac{\alpha_{s}(\mu)}{\alpha_{s}(m_b)}\right]^{\frac{12}{23}} \, ,\nonumber\\
\alpha_s(\mu)&=&\frac{1}{b_0t}\left[1-\frac{b_1}{b_0^2}\frac{\log t}{t} +\frac{b_1^2(\log^2{t}-\log{t}-1)+b_0b_2}{b_0^4t^2}\right]\, ,
\end{eqnarray}
  where $t=\log \frac{\mu^2}{\Lambda^2}$, $b_0=\frac{33-2n_f}{12\pi}$, $b_1=\frac{153-19n_f}{24\pi^2}$, $b_2=\frac{2857-\frac{5033}{9}n_f+\frac{325}{27}n_f^2}{128\pi^3}$,  $\Lambda=213\,\rm{MeV}$, $296\,\rm{MeV}$  and  $339\,\rm{MeV}$ for the flavors  $n_f=5$, $4$ and $3$, respectively  \cite{PDG}.

 The values of the thresholds are $2M_{\eta_c}=5966.8\,\rm{MeV}$, $2M_{J/\psi}=6193.8\,\rm{MeV}$, $2M_{\eta_b}=18798.0\,\rm{MeV}$, $2M_{\Upsilon}=18920.6\,\rm{MeV}$ from the Particle Data Group \cite{PDG}. The masses of the $0^{++}$ and $2^{++}$ $QQ\bar{Q}\bar{Q}$ tetraquark states from the  phenomenological quark models lie above or below those thresholds \cite{Silvestre-1986,Lloyd-2004,Barnea-2006,Bai-2016,Heupel-2012,Berezhnoy-2012,Rosner-2016,Wu-2016,Chen-2016}. In Ref.\cite{Wang-BC},
 we study the vector and axialvector $B_c$ mesons with the QCD sum rules
and obtain the masses $M_{B_{c}^*}=6.337\pm0.052\,\rm{GeV}$ and $M_{B_{c1}}=6.730\pm0.061\,\rm{GeV}$ at the typical energy scale $\mu=2\,\rm{GeV}$. The $B_c$ mesons have two heavy quarks, and the mass $M_{B_{c}^*}=6.337\pm0.052\,\rm{GeV}$ lies slightly above the threshold  $2M_{J/\psi}=6193.8\,\rm{MeV}$, so we expect the ideal energy scale to extract masses of the $cc\bar{c}\bar{c}$ tetraquark states  from the QCD sum rules is about $\mu=2\,\rm{GeV}$, it is indeed the case.

In Fig.1, we plot the masses of the  $cc\bar{c}\bar{c}$ tetraquark states with variations of the energy scales and Borel parameters for the threshold parameters $s_S^0=42\,\rm{GeV}^2$ and $S_T^0=44\,\rm{GeV}^2$. From the figure, we can see that the predicted masses decrease  monotonously and slowly with increase of the energy scales, the QCD sum rules  are stable with variations of the Borel parameters at the energy scales $1.2\,{\rm{GeV}}<\mu<2.2\,{\rm GeV}$. At the energy scale $\mu=2.0\,\rm{GeV}$, the relation $S_{S/T}^0=\left(M_{S/T}+0.5\,\rm{GeV}\right)^2$ is satisfied, naively, we expect that the energy gap between the ground states and the first radial excited states ia about $0.5\,\rm{GeV}$. In the QCD sum rules, we usually take the continuum threshold parameters as $\sqrt{s_0}=M_{gr}+(0.4\sim 0.6)\,\rm{GeV}$ for the conventional mesons, where the $gr$ denotes the ground states.
  Experimentally, the energy gaps $M_{\psi^\prime}-M_{J/\psi}=589\,\rm{MeV}$ and    $M_{\eta_c^\prime}-M_{\eta_c}=656\,\rm{MeV}$ from the Particle Data Group \cite{PDG}.
 Now we revisit the mass gaps of the tetraquark states.
 In Ref.\cite{Wang4430}, we tentatively assign the $Z_c(3900)$ and $Z(4430)$  to be  the ground state and the first radial excited state of the axial-vector  tetraquark states with $J^{PC}=1^{+-}$, respectively, and reproduce the experimental values of the masses with the QCD sum rules.
 In Ref.\cite{Wang4500}, we tentatively assign the  $X(3915)$ and $X(4500)$ to be the ground state and the first radial excited state of the scalar $cs\bar{c}\bar{s}$ tetraquark states with $J^{PC}=0^{++}$, respectively, and reproduce the experimental values of the masses with the QCD sum rules.
  The mass gaps are $M_{Z(4430)}-M_{Z_c(3900)}=576\,\rm{MeV}$ and $M_{X(4500)}-M_{X(3915)}=588\,\rm{MeV}$, which also satisfy the relation $\sqrt{s_0}=M_{gr}+(0.4\sim 0.6)\,\rm{GeV}$, if only the ground states are taken into account in the QCD sum rules.
 In this article, we take the relation $\sqrt{s_0}=M_{gr}+(0.4\sim 0.6)\,\rm{GeV}$ as a constraint, and search for the optimal values of the $s_0$ to reproduce the $M_{gr}$.  At the regions $S_{S/T}^0\leq \left(M_{S/T}+0.6\,\rm{GeV}\right)^2$, the contributions of the excited states are expected not to be included in.

 \begin{figure}
 \centering
 \includegraphics[totalheight=5cm,width=7cm]{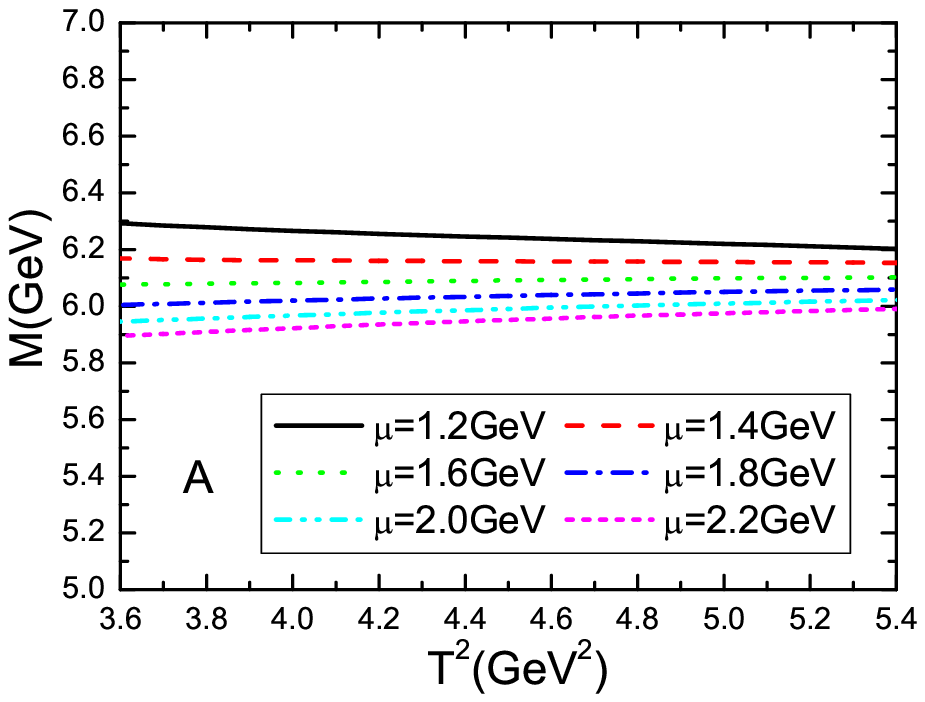}
 \includegraphics[totalheight=5cm,width=7cm]{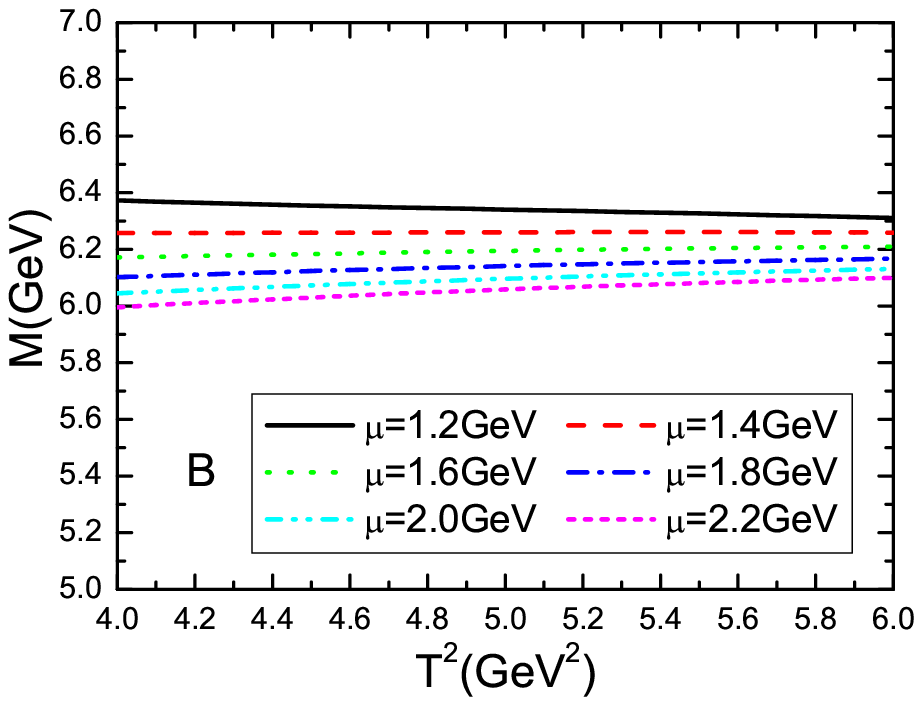}
   \caption{ The masses  of the $cc\bar{c}\bar{c}$ tetraquark states  with variations of the energy scales and Borel parameters,  where the $A$ and $B$ denote the $cc\bar{c}\bar{c}(0^{++})$ and $cc\bar{c}\bar{c}(2^{++})$, respectively.  }
\end{figure}

In Fig.2, we plot the masses of the  $bb\bar{b}\bar{b}$ tetraquark states with variations of the energy scales and Borel parameters for the threshold parameters $s_S^0=374\,\rm{GeV}^2$ and $S_T^0=375\,\rm{GeV}^2$. From the figure, we can see that the predicted masses also decrease  monotonously and slowly  with increase of the energy scales, the QCD sum rules  are stable with variations of the Borel parameters at the energy scales $2.5\,{\rm{GeV}}<\mu<3.3\,{\rm GeV}$. At the energy scale $\mu=3.1\,\rm{GeV}$, the relation $S_{S/T}^0=\left(M_{S/T}+0.5\,\rm{GeV}\right)^2$ is satisfied. For the conventional mesons, the energy gaps $M_{\Upsilon^\prime}-M_{\Upsilon}=563\,\rm{MeV}$ and    $M_{\eta_b^\prime}-M_{\eta_b}=600\,\rm{MeV}$ from the Particle Data Group \cite{PDG}. It is also reasonable to   choose the relation $S_{S/T}^0\leq \left(M_{S/T}+0.6\,\rm{GeV}\right)^2$.
\begin{figure}
 \centering
  \includegraphics[totalheight=5cm,width=7cm]{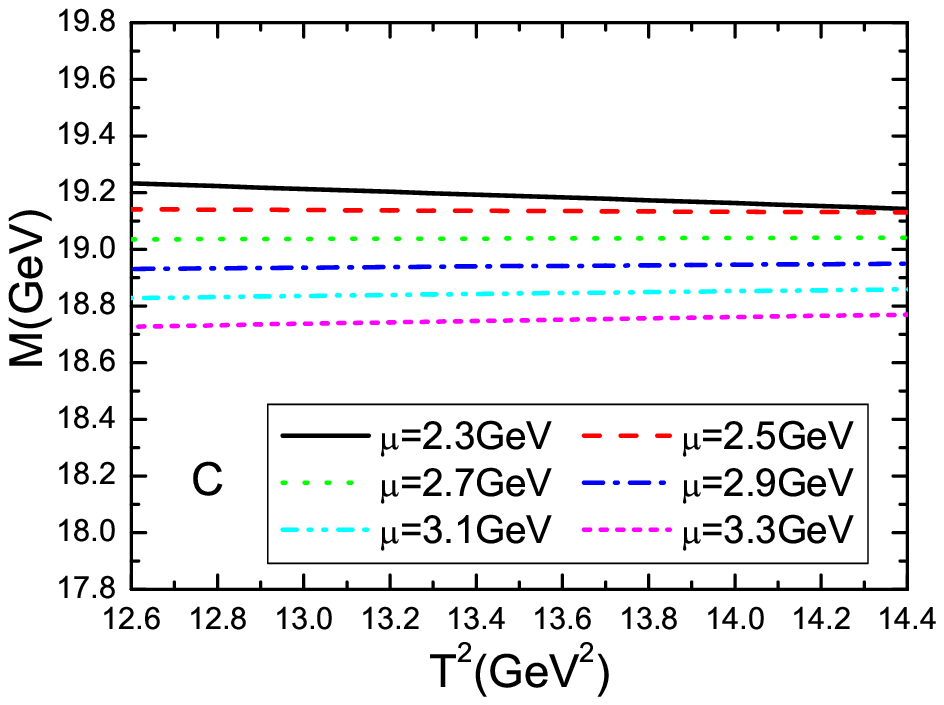}
 \includegraphics[totalheight=5cm,width=7cm]{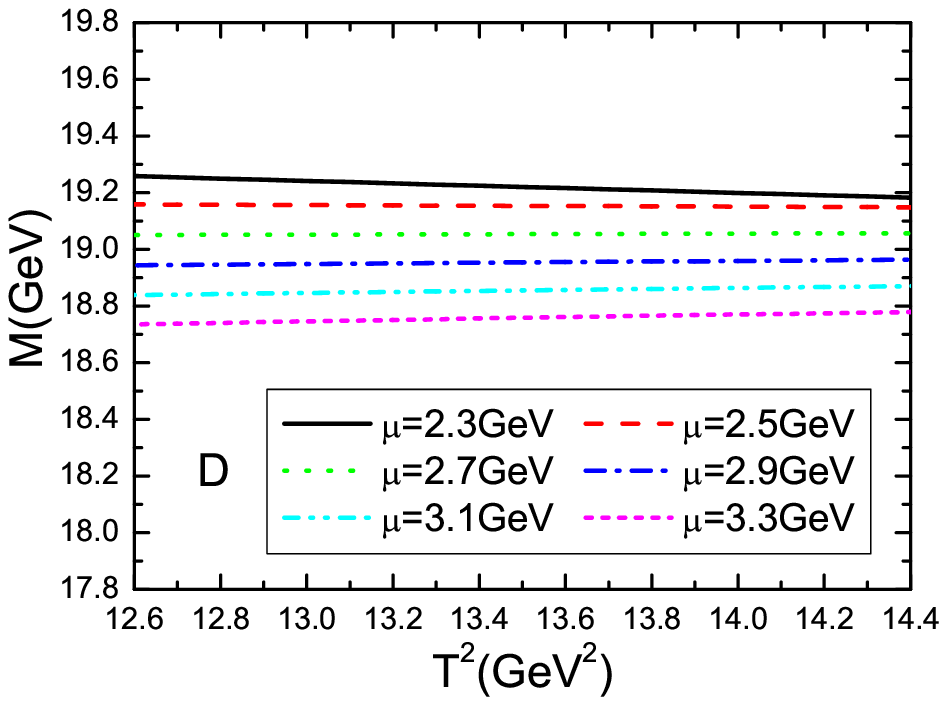}
        \caption{ The masses  of the $bb\bar{b}\bar{b}$ tetraquark states  with variations of the energy scales and Borel parameters,  where the $C$ and $D$ denote the $bb\bar{b}\bar{b}(0^{++})$ and $bb\bar{b}\bar{b}(2^{++})$, respectively.  }
\end{figure}

 We search for the  Borel parameters $T^2$ and continuum threshold
parameters $s_0$  to satisfy the  two  criteria of the QCD sum rules:  pole dominance at the phenomenological side and
convergence of the operator product expansion at the QCD side. Furthermore, we take the relation $\sqrt{S_{S/T}^0}= M_{S/T}+(0.4\sim0.6)\,\rm{GeV}$ as an additional constraint to obey.
The resulting Borel parameters, continuum threshold parameters, energy scales, pole contributions are shown explicitly in Table 1. From the Table, we can see that the pole dominance at the phenomenological side is well satisfied. In the Borel windows, the dominant contributions come from the perturbative terms, the  contributions of the gluon condensate are about $-10\%$, the operator product expansion is well convergent. Now the two    criteria of the QCD sum rules  are all satisfied, we expect to make reasonable predictions.

In Ref.\cite{WangD3K}, we tentatively assign the $D_{s3}^*(2860)$ to be  a D-wave $c\bar{s}$  meson, and  study the mass and decay constant of the $D_{s3}^*(2860)$ with the QCD sum rules by calculating the contributions of the vacuum condensates up to dimension-6 in the operator product expansion. In calculations, we observe that only the perturbative term, gluon condensate and three-gluon condensate have contributions. At the Borel window, the contributions are about $(107-109)\%$, $-(7-9)\%$ and $\ll1\%$, respectively, see the first diagram in Fig.3 \cite{WangD3K}, the three-gluon condensate can be neglected safely. In the present case,  the  contributions of the gluon condensate are about $-10\%$, just like in the case of the QCD sum rules for the $D_{s3}^*(2860)$, so neglecting the three-gluon condensate cannot impair the predictive ability.  As the dominant contributions come from the perturbative terms, perturbative $\mathcal{O}(\alpha_s)$ corrections amount to multiplying the pertubative terms by a factor $\kappa$, which can be absorbed into the pole residues and cannot impair the predicted masses remarkably.

\begin{table}
\begin{center}
\begin{tabular}{|c|c|c|c|c|c|c|c|}\hline\hline
                            &$T^2(\rm{GeV}^2)$  &$s_0(\rm{GeV}^2)$ &$\mu(\rm{GeV})$ &pole         &$M_{X}(\rm{GeV})$    &$\lambda_{X}(\rm{GeV}^5)$ \\ \hline
$cc\bar{c}\bar{c}(0^{++})$  &$4.2-4.6$          &$42\pm1$          &$2.0$           &$(46-62)\%$  &$5.99\pm0.08$        &$(3.72\pm0.54)\times10^{-1}$  \\
$cc\bar{c}\bar{c}(2^{++})$  &$4.6-5.2$          &$44\pm1$          &$2.0$           &$(46-63)\%$  &$6.09\pm0.08$        &$(3.36\pm0.45)\times10^{-1}$  \\
$bb\bar{b}\bar{b}(0^{++})$  &$13.0-13.6$        &$374\pm3$         &$3.1$           &$(49-61)\%$  &$18.84\pm0.09$       &$6.79\pm1.27$  \\
$bb\bar{b}\bar{b}(2^{++})$  &$13.0-13.6$        &$375\pm3$         &$3.1$           &$(51-63)\%$  &$18.85\pm0.09$       &$5.55\pm1.00$  \\  \hline
\end{tabular}
\end{center}
\caption{ The Borel parameters, continuum threshold parameters, energy scales,  pole contributions, masses and pole residues of the tetraquark states. }
\end{table}

We take into account all uncertainties of the input parameters, and obtain the values of the ground state masses and pole residues, which are also shown explicitly in Table 1 and Figs.3-4. From Table 1, we can see that  the additional constraint is also satisfied. In Figs.3-4, we plot the masses and pole residues with variations
of the Borel parameters at much larger intervals   than the  Borel windows shown in Table 1.
From Figs.3-4, we can see that the predicted masses and pole residues are rather stable with variations of the Borel parameters, the uncertainties originate from the Borel parameters in the Borel windows are very small.

From Table 1, we can see that the mass splitting between the $0^{++}$ and $2^{++}$ $bb\bar{b}\bar{b}$ tetraquark states is much smaller than that for the $cc\bar{c}\bar{c}$ tetraquark states.
The heavy quark effective Lagrangian  can be written as
\begin{eqnarray}
  {\cal L}  &=&\bar h_viv\cdot Dh_v+{1\over2m_Q}  \left[\bar h_v(iD_\perp)^2h_v
  +{g_s\over2}\,\bar h_v\sigma^{\alpha\beta}  G_{\alpha\beta}h_v\right]  +\dots\,,
\end{eqnarray}
where $D^\mu_\perp=D^\mu-v^\mu v\cdot D$, the $D_\mu$ is the
covariant derivative, and the $h_v$ is the effective heavy quark field.
The heavy quark spin symmetry breaking terms appear at the order  $\mathcal{O}(1/m_Q)$ \cite{RevWise}. The $\overline{MS}$ masses are $m_{c}(m_c)=(1.275\pm0.025)\,\rm{GeV}$ and $m_{b}(m_b)=(4.18\pm0.03)\,\rm{GeV}$ from the Particle Data Group \cite{PDG}, the heavy quark spin symmetry breaking effects in the $c$-quark systems are much larger than that  in the $b$-quark systems.

\begin{figure}
 \centering
 \includegraphics[totalheight=5cm,width=7cm]{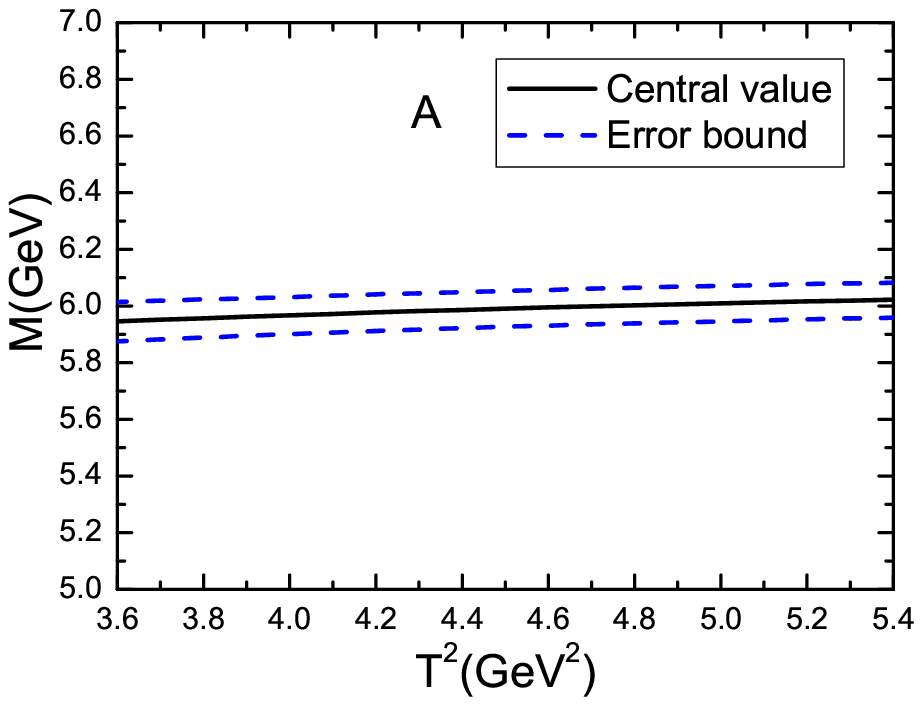}
 \includegraphics[totalheight=5cm,width=7cm]{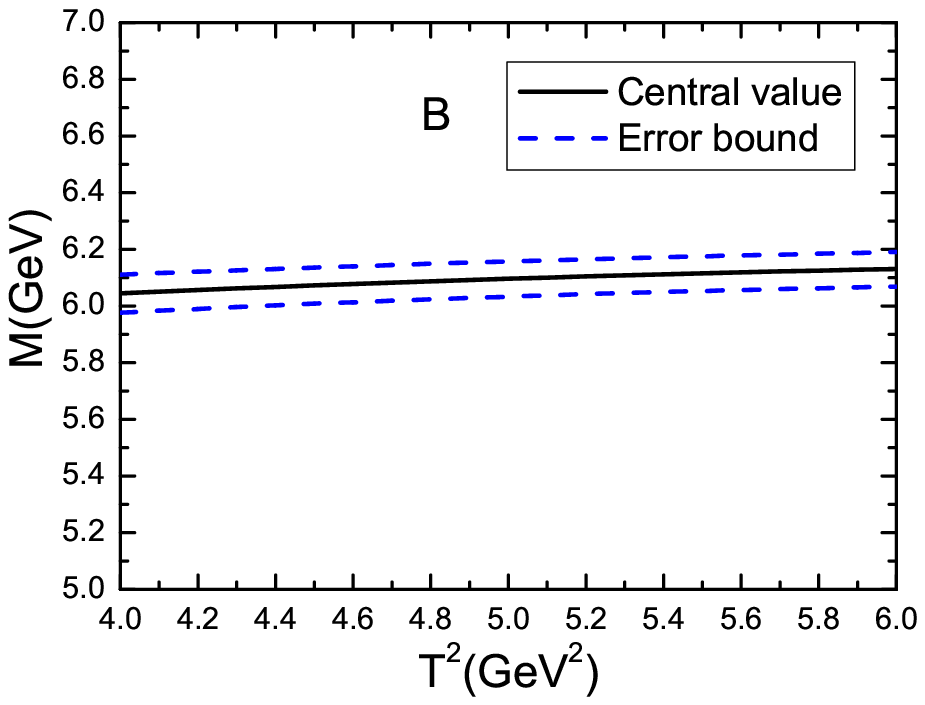}
 \includegraphics[totalheight=5cm,width=7cm]{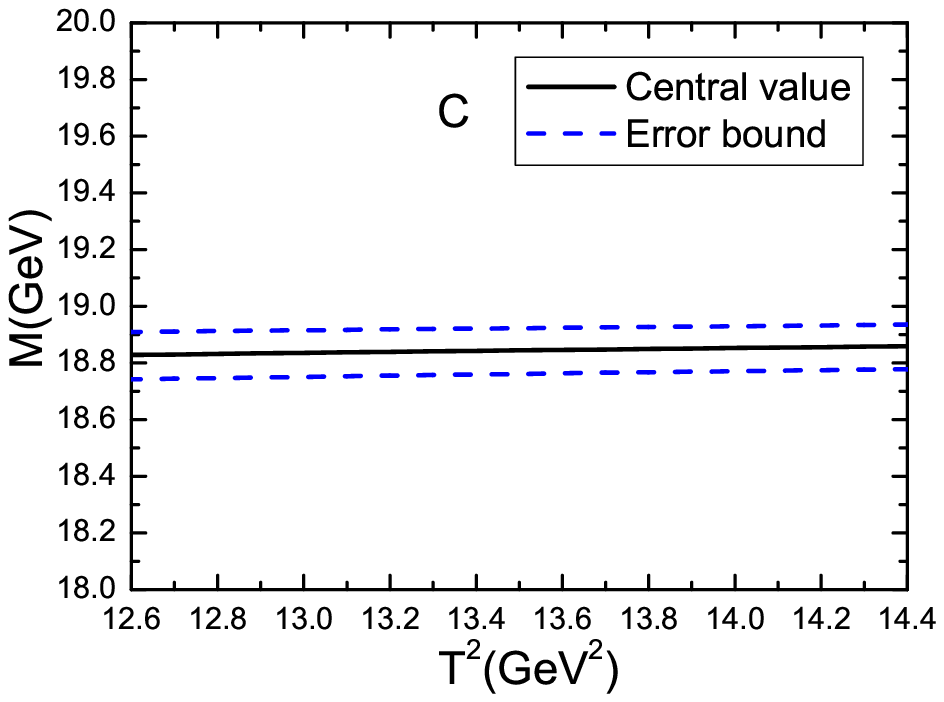}
 \includegraphics[totalheight=5cm,width=7cm]{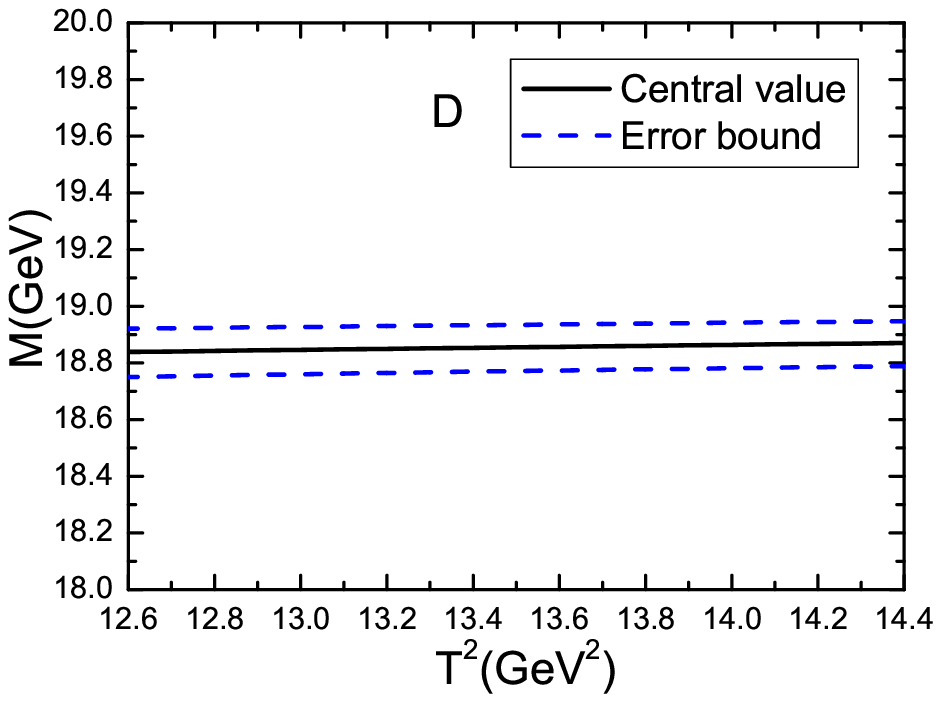}
        \caption{ The masses   of the tetraquark states  with variations of the Borel parameters $T^2$,  where the $A$, $B$, $C$ and $D$ denote the $cc\bar{c}\bar{c}(0^{++})$, $cc\bar{c}\bar{c}(2^{++})$, $bb\bar{b}\bar{b}(0^{++})$ and $bb\bar{b}\bar{b}(2^{++})$, respectively.  }
\end{figure}

\begin{figure}
 \centering
 \includegraphics[totalheight=5cm,width=7cm]{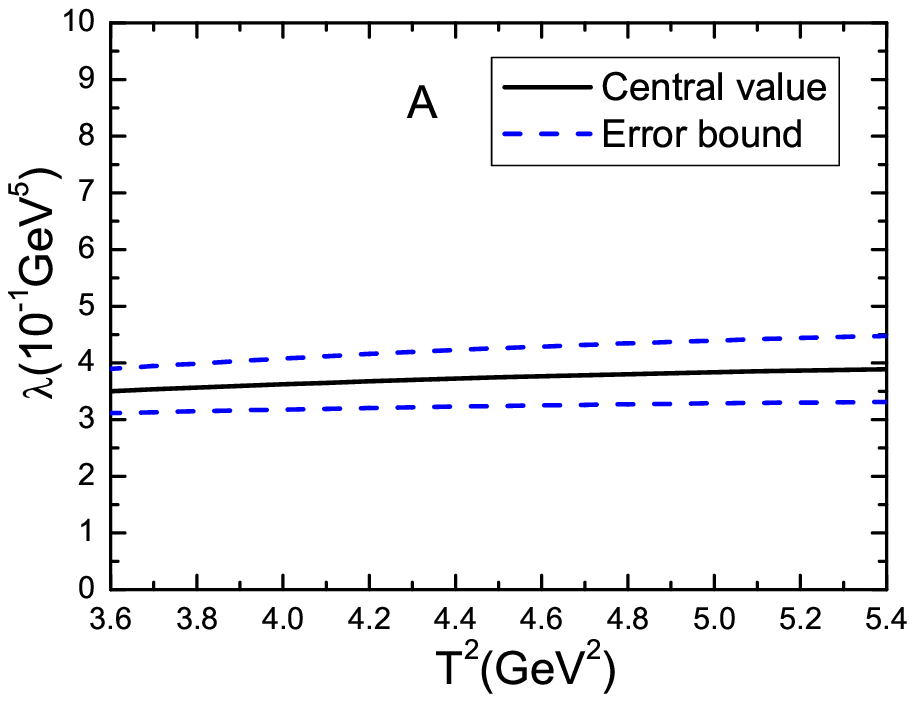}
 \includegraphics[totalheight=5cm,width=7cm]{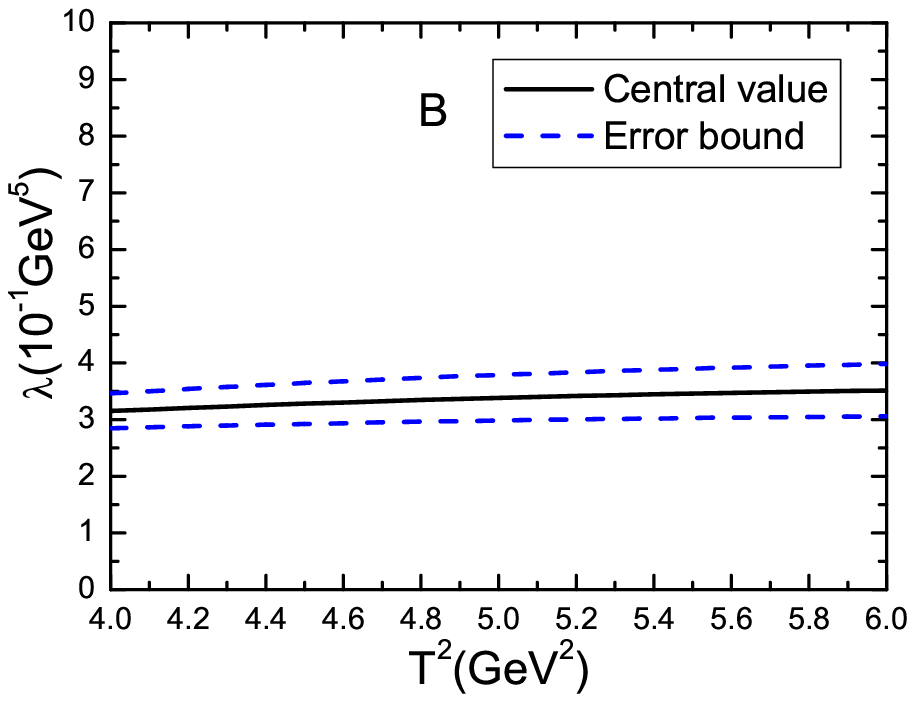}
 \includegraphics[totalheight=5cm,width=7cm]{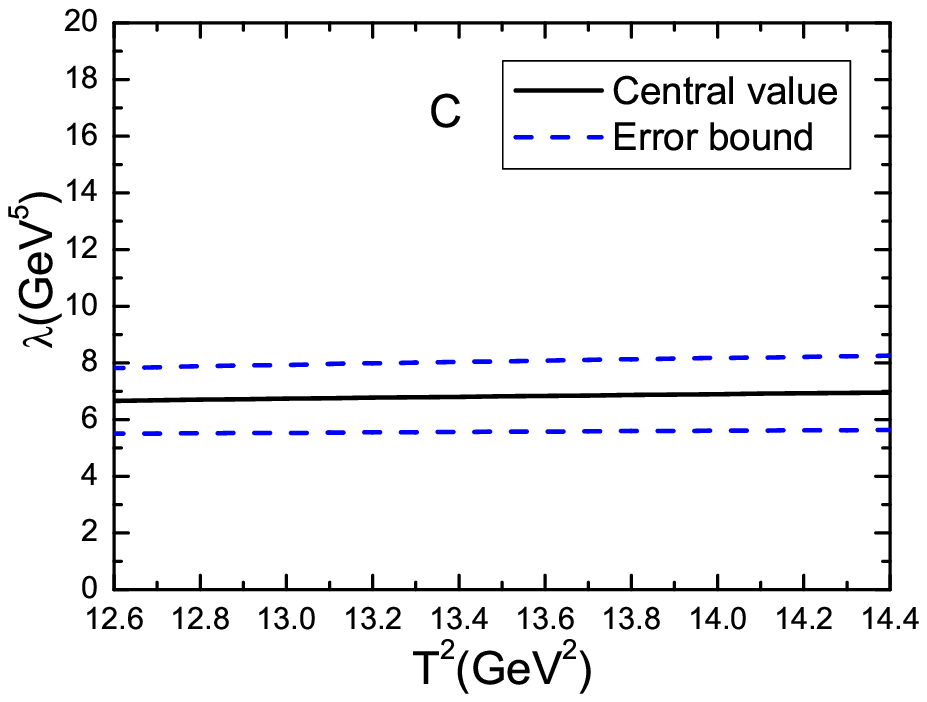}
 \includegraphics[totalheight=5cm,width=7cm]{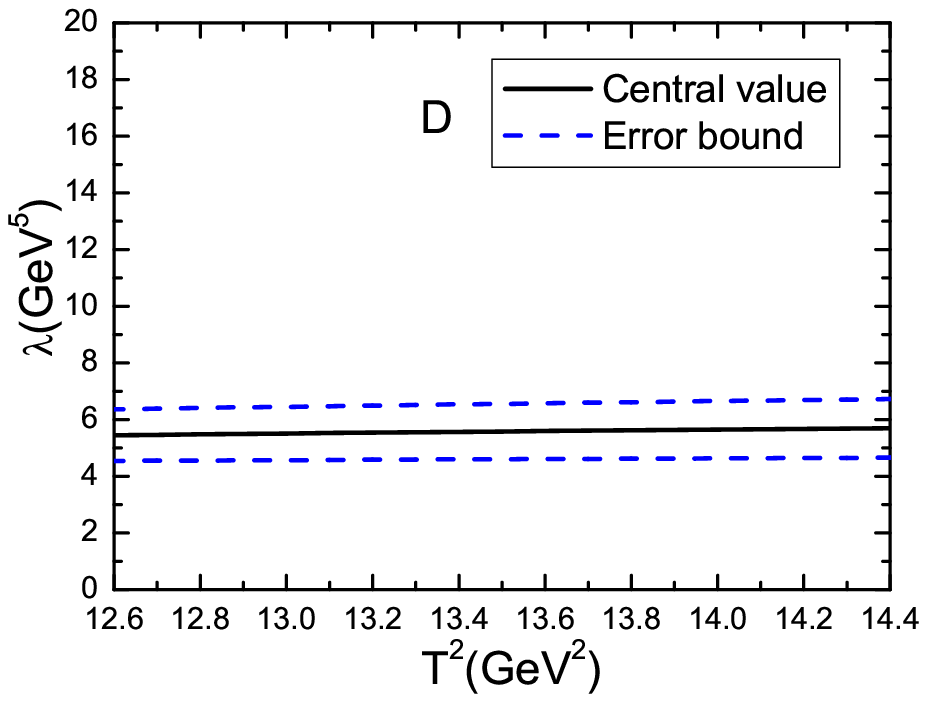}
        \caption{ The pole residues   of the tetraquark states  with variations of the Borel parameters $T^2$,  where the $A$, $B$, $C$ and $D$ denote the $cc\bar{c}\bar{c}(0^{++})$, $cc\bar{c}\bar{c}(2^{++})$, $bb\bar{b}\bar{b}(0^{++})$ and $bb\bar{b}\bar{b}(2^{++})$, respectively.  }
\end{figure}

In 2002,  the SELEX collaboration reported   the first observation
of a signal for the double-charm baryon state  $ \Xi_{cc}^+$ in
the decay mode $\Xi_{cc}^+\rightarrow\Lambda_c^+K^-\pi^+$
\cite{SELEX2002}, and confirmed later by the same collaboration in
the decay mode $\Xi_{cc}^+\rightarrow pD^+K^- $ with the measured
mass $M_{\Xi}=(3518.9 \pm 0.9) \,\rm{ MeV }$ \cite{SELEX2004}.
In Ref.\cite{Wang-Xcc}, we study the  ${1\over 2}^+$ doubly heavy baryon
states $\Omega_{QQ}$ and $\Xi_{QQ}$  by subtracting the
contributions from the corresponding ${1\over 2}^-$ doubly heavy
baryon states with the QCD sum rules, and obtain the value    $M_{\Xi_{cc}}=3.57\pm0.14\,\rm{GeV}$.
If exciting  additional quark-antiquark pair $q\bar{q}$ with $J^{PC}=0^{++}$ costs energy about $1\,\rm{GeV}$, then $M_{X(cc\bar{c}\bar{c},0^{++}/2^{++})}\approx 2M_{\Xi_{cc}}-1\,\rm{GeV}=6.14\pm0.14\,\rm{GeV}$, which is consistent with the present prediction.

The predicted masses are
\begin{eqnarray}
M_{X(cc\bar{c}\bar{c},0^{++})}  &=&5.91-6.07\,\rm{GeV} \, ,    \nonumber\\
M_{X(cc\bar{c}\bar{c},2^{++})}  &=&6.01-6.17\,\rm{GeV} \, ,    \nonumber\\
M_{X(bb\bar{b}\bar{b},0^{++})}  &=&18.75-18.93\,\rm{GeV} \, ,    \nonumber\\
M_{X(bb\bar{b}\bar{b},2^{++})}  &=&18.76-18.94\,\rm{GeV} \, .
\end{eqnarray}
The decays
\begin{eqnarray}
X(cc\bar{c}\bar{c},0^{++}/2^{++}) &\to& \eta_c \eta_c \to \gamma\gamma\gamma\gamma\, ,\nonumber\\
X(bb\bar{b}\bar{b},0^{++}/2^{++}) &\to& \eta_b \eta_b \to \gamma\gamma\gamma\gamma \, ,
\end{eqnarray}
are kinematically allowed, but the available spaces are small.
The decays
\begin{eqnarray}
X(cc\bar{c}\bar{c},0^{++}/2^{++}) &\to& J/\psi J/\psi \, ,\nonumber\\
X(bb\bar{b}\bar{b},0^{++}/2^{++}) &\to& \Upsilon \Upsilon \, ,
\end{eqnarray}
are kinematically forbidden, but the decays
 \begin{eqnarray}
X(cc\bar{c}\bar{c},0^{++}/2^{++}) &\to& J/\psi {J/\psi}^* \to \mu^+\mu^- \mu^+\mu^- \, ,\nonumber\\
X(bb\bar{b}\bar{b},0^{++}/2^{++}) &\to& \Upsilon \Upsilon^* \to \mu^+\mu^- \mu^+\mu^-\, ,
\end{eqnarray}
can take place,
we can search for the $X(cc\bar{c}\bar{c},0^{++}/2^{++})$ and $X(bb\bar{b}\bar{b},0^{++}/2^{++}) $ in the mass spectrum of the $\mu^+\mu^- \mu^+\mu^-$ in the future.

In the following, we perform Fierz re-arrangement  to the tensor current $J_{\mu\nu}$ and scalar current    $J$ both in the color and Dirac-spinor  spaces to   obtain the results,
\begin{eqnarray}
J_{\mu\nu} &=&\frac{1}{\sqrt{2}} \Big\{\, \bar{Q}\gamma_\mu\gamma_5Q\, \bar{Q}\gamma_\nu\gamma_5Q+\bar{Q}\gamma_\nu\gamma_5Q\, \bar{Q}\gamma_\mu\gamma_5Q -\bar{Q}\gamma_\mu Q\, \bar{Q}\gamma_\nu Q-\bar{Q}\gamma_\nu Q\, \bar{Q}\gamma_\mu Q  \nonumber\\
  &&+g^{\alpha\beta}\left(\bar{Q}\sigma_{\mu\alpha}Q\, \bar{Q}\sigma_{\nu\beta}Q+\bar{Q}\sigma_{\nu\alpha}Q\, \bar{Q}\sigma_{\mu\beta}Q\right) +g_{\mu\nu}\Big( \bar{Q}Q\,\bar{Q}Q+\bar{Q}i\gamma_5Q\,\bar{Q}i\gamma_5Q\nonumber\\
 && +\bar{Q}\gamma_{\alpha} Q\,\bar{Q}\gamma^{\alpha}Q-\bar{Q}\gamma_{\alpha}\gamma_5 Q\,\bar{Q}\gamma^{\alpha}\gamma_5Q-\frac{1}{2}\bar{Q}\sigma_{\alpha\beta} Q\,\bar{Q}\sigma^{\alpha\beta}Q\Big) \Big\} \, ,
\end{eqnarray}
\begin{eqnarray}
J &=&  2\bar{Q}Q\,\bar{Q}Q+2\bar{Q}i\gamma_5Q\,\bar{Q}i\gamma_5Q+\bar{Q}\gamma_{\alpha} Q\,\bar{Q}\gamma^{\alpha}Q-\bar{Q}\gamma_{\alpha}\gamma_5 Q\,\bar{Q}\gamma^{\alpha}\gamma_5Q \, .
\end{eqnarray}
Now we can see that the diquark-antidiquark type current can be  changed  to a current as a special superposition of   color  singlet-singlet type currents, which couple potentially to the meson-meson pairs. The
diquark-antidiquark type tetraquark state can be taken as a special superposition of a series of  meson-meson pairs, and embodies  the net effects. The decays to its components (meson-meson pairs) are Okubo-Zweig-Iizuka super-allowed,  but the re-arrangements in the color-space are highly non-trivial.

We take the current $J$ as an example to illustrate that the scalar tetraquark state can embody the net effects of all the meson-meson pairs.
At the phenomenological side, we can insert  a complete set of intermediate hadronic states with
the same quantum numbers as the current operator $J(x)$ into the
correlation function $\Pi(p)$ to obtain the hadronic representation
\cite{SVZ79,Reinders85}. After isolating the lowest meson-meson pairs, we get the following result,
\begin{eqnarray}
\Pi (p) &=&\frac{i}{(2\pi)^4}\int d^4q \frac{i}{q^2-M_{\eta_c}^2}\frac{i}{(q+p)^2-M_{\eta_c}^2} \left\{\frac{f_{\eta_c}^4q^4(q+p)^4}{4m_c^4} +f_{\eta_c}^4 \left[q \cdot (q+p)\right]^2 \right. \nonumber\\
&&\left.+\frac{f_{\eta_c}^4q^2(q+p)^2\,q \cdot (q+p) }{m_c^2}\right\} \nonumber\\
&&+\frac{i}{(2\pi)^4}\int d^4q \frac{i}{q^2-M_{J/\psi}^2}\frac{i}{(q+p)^2-M_{J/\psi}^2} f_{J/\psi}^4 q^2(q+p)^2 \left[g_{\mu\nu}-\frac{q_\mu q_\nu}{q^2} \right]  \nonumber\\
&& \left[g^{\mu\nu}-\frac{(q+p)^\mu (q+p)^\nu}{(q+p)^2} \right]   \nonumber\\
&&+\frac{i}{(2\pi)^4}\int d^4q \frac{i}{q^2-M_{\chi_{c0}}^2}\frac{i}{(q+p)^2-M_{\chi_{c0}}^2} \left\{4f_{\chi_{c0}}^4  q^2(q+p)^2  + f_{\chi_{c0}}^4\left[q \cdot (q+p) \right]^2 \right. \nonumber\\
&&\left.+4f_{\chi_{c0}}^4  \sqrt{q^2(q+p)^2} \, q \cdot (q+p) \right\} \nonumber\\
&&+\frac{i}{(2\pi)^4}\int d^4q \frac{i}{q^2-M_{\chi_{c1}}^2}\frac{i}{(q+p)^2-M_{\chi_{c1}}^2} f_{\chi_{c1}}^4 q^2(q+p)^2 \left[g_{\mu\nu}-\frac{q_\mu q_\nu}{q^2} \right]  \nonumber\\
&& \left[g^{\mu\nu}-\frac{(q+p)^\mu (q+p)^\nu}{(q+p)^2} \right] +\cdots  \, ,
\end{eqnarray}
where the decay constants $f_{\eta_c}$, $f_{J/\psi}$, $f_{\chi_{c0}}$ and $f_{\chi_{c1}}$ are defined by
\begin{eqnarray}
 \langle 0|J(0)|\eta_c(q)\eta_c(q+p)\rangle&=&2\frac{f_{\eta_c}q^2}{2m_c} \frac{f_{\eta_c}(q+p)^2}{2m_c}+f_{\eta_c}^2 \,q \cdot (q+p)  \, , \nonumber\\
 \langle 0|J(0)|J/\psi(q)J/\psi(q+p)\rangle&=&f_{J/\psi}^2 \sqrt{q^2(q+p)^2}\, \varepsilon_\alpha \varepsilon^\alpha  \, , \nonumber\\
 \langle 0|J(0)|\chi_{c0}(q)\chi_{c0}(q+p)\rangle&=&2f_{\chi_{c0}}^2 \sqrt{q^2(q+p)^2} + f_{\chi_{c0}}^2\,q \cdot (q+p) \, , \nonumber\\
  \langle 0|J(0)|\chi_{c1}(q)\chi_{c1}(q+p)\rangle&=&f_{\chi_{c1}}^2 \sqrt{q^2(q+p)^2}\, \varepsilon_\alpha \varepsilon^\alpha  \, ,
\end{eqnarray}
the $\varepsilon_{\mu}$ are the  polarization vectors of the $J/\psi$ and $\chi_{c1}$.

We can rewrite the correlation function $\Pi(p)$ into  the following form through dispersion relation,
\begin{eqnarray}
\Pi (p) &=&\frac{f_{\eta_c}^4}{16\pi^2}\int_{4M_{\eta_c}^2}^{s_0}ds \frac{1}{s-p^2} \left\{\frac{M_{\eta_c}^8}{4m_c^4} + \left(\frac{s}{2}-M_{\eta_c}^2\right)^2-\frac{M_{\eta_c}^4}{m_c^2} \left(\frac{s}{2}-M_{\eta_c}^2\right) \right\}\sqrt{1-\frac{4M_{\eta_c}^2}{s}} \nonumber\\
&&+\frac{f_{J/\psi}^4}{16\pi^2}\int_{4M_{J/\psi}^2}^{s_0}ds \frac{1}{s-p^2}  M_{J/\psi}^4 \left\{2 +\frac{ \left(\frac{s}{2}-M_{J/\psi}^2\right)^2}{M_{J/\psi}^4} \right\} \sqrt{1-\frac{4M_{J/\psi}^2}{s}}  \nonumber\\
&&+\frac{f_{\chi_{c0}}^4}{16\pi^2}\int_{4M_{\chi_{c0}}^2}^{s_0}ds \frac{1}{s-p^2} \left\{4M_{\chi_{c0}}^4 + \left(\frac{s}{2}-M_{\chi_{c0}}^2\right)^2
-4M_{\chi_{c0}}^2 \left(\frac{s}{2}-M_{\chi_{c0}}^2\right)  \right\}\sqrt{1-\frac{4M_{\chi_{c0}}^2}{s}} \nonumber\\
&&+\frac{f_{\chi_{c1}}^4}{16\pi^2}\int_{4M_{\chi_{c1}}^2}^{s_0}ds \frac{1}{s-p^2}  M_{\chi_{c1}}^4 \left\{2 +\frac{ \left(\frac{s}{2}-M_{\chi_{c1}}^2\right)^2}{M_{\chi_{c1}}^4} \right\} \sqrt{1-\frac{4M_{\chi_{c1}}^2}{s}}  +\cdots\, .
\end{eqnarray}

In this article, we choose the value $s_0<4M_{\chi_{c0}}^2,\,4M_{\chi_{c1}}^2$, the meson pairs $\chi_{c0}\chi_{c0}$ and $\chi_{c1}\chi_{c1}$  have no contributions, the QCD sum rules can be written as
\begin{eqnarray}
&&\frac{f_{\eta_c}^4}{16\pi^2}\int_{4M_{\eta_c}^2}^{s_0}ds  \left\{\frac{M_{\eta_c}^8}{4m_c^4} + \left(\frac{s}{2}-M_{\eta_c}^2\right)^2-\frac{M_{\eta_c}^4}{m_c^2} \left(\frac{s}{2}-M_{\eta_c}^2\right) \right\}\sqrt{1-\frac{4M_{\eta_c}^2}{s}} \,s\, \exp\left(-\frac{s}{T^2}\right)\nonumber\\
&&+\frac{f_{J/\psi}^4}{16\pi^2}\int_{4M_{J/\psi}^2}^{s_0}ds   M_{J/\psi}^4 \left\{2 +\frac{ \left(\frac{s}{2}-M_{J/\psi}^2\right)^2}{M_{J/\psi}^4} \right\} \sqrt{1-\frac{4M_{J/\psi}^2}{s}} \,s\, \exp\left(-\frac{s}{T^2}\right) \nonumber\\
&&= \kappa^2 \,\int_{16m_c^2}^{s_0} ds \int_{z_i}^{z_f}dz \int_{t_i}^{t_f}dt \int_{r_i}^{r_f}dr\, \rho_S(s,z,t,r)  \,s\, \exp\left(-\frac{s}{T^2}\right) \, ,
\end{eqnarray}

\begin{eqnarray}
&&\frac{f_{\eta_c}^4}{16\pi^2}\int_{4M_{\eta_c}^2}^{s_0}ds  \left\{\frac{M_{\eta_c}^8}{4m_c^4} + \left(\frac{s}{2}-M_{\eta_c}^2\right)^2-\frac{M_{\eta_c}^4}{m_c^2} \left(\frac{s}{2}-M_{\eta_c}^2\right) \right\}\sqrt{1-\frac{4M_{\eta_c}^2}{s}}\, \exp\left(-\frac{s}{T^2}\right)\nonumber\\
&&+\frac{f_{J/\psi}^4}{16\pi^2}\int_{4M_{J/\psi}^2}^{s_0}ds   M_{J/\psi}^4 \left\{2 +\frac{ \left(\frac{s}{2}-M_{J/\psi}^2\right)^2}{M_{J/\psi}^4} \right\} \sqrt{1-\frac{4M_{J/\psi}^2}{s}}\, \exp\left(-\frac{s}{T^2}\right) \nonumber\\
&&=\kappa^2 \, \int_{16m_c^2}^{s_0} ds \int_{z_i}^{z_f}dz \int_{t_i}^{t_f}dt \int_{r_i}^{r_f}dr\, \rho_S(s,z,t,r)  \, \exp\left(-\frac{s}{T^2}\right) \, ,
\end{eqnarray}
where we introduce a coefficient $\kappa$, if $\kappa=1$, the QCD sum rules can be saturated by the meson pairs $\eta_c\eta_c$ and $J/\psi J/\psi$.

We choose the input parameters as $M_{\eta_c}=2.9834\,\rm{GeV}$, $M_{J/\psi}=3.0969\,\rm{GeV}$ \cite{PDG}, $f_{\eta_c}=0.387\,\rm{GeV}$,  $f_{J/\psi}=0.418\,\rm{GeV}$ \cite{feta-fpsi}, $s_0=42\,\rm{GeV}^2$. In Fig.5, we plot the coefficient $\kappa$ comes  from Eq.(28) with variation of the Borel parameter $T^2$, from the figure, we can see that the values of the $\kappa$ are rather stable with variation of the Borel parameter. Now we choose the special value $T^2=4.4\,\rm{GeV}^2$, and plot the coefficient $\kappa$  with variation of the energy scale $\mu$ in Fig.6. From the figure, we can see that the coefficient $\kappa$ decreases monotonously and quickly with increase of the energy scale $\mu$ at the region $\mu\leq 1.6\,\rm{GeV}$. At the vicinity of the energy scale $\mu=1.5\,\rm{GeV}$,  $\kappa\approx 1$, however, the
reliable  QCD sum rules do not depend heavily on the energy scale $\mu$. So the QCD sum rules cannot be saturated by the  meson pairs $\eta_c\eta_c$ and $J/\psi J/\psi$.

Now we saturate the QCD sum rules by the meson pairs $\eta_c\eta_c$, $J/\psi J/\psi$ plus a scalar tetraquark state $X(cc\bar{c}\bar{c},0^{++})$ at the phenomenological side,
\begin{eqnarray}
&&\lambda^2_{X}\,M^2_{X}\, \exp\left(-\frac{M^2_{X}}{T^2}\right) \nonumber\\
&&+\frac{f_{\eta_c}^4}{16\pi^2}\int_{4M_{\eta_c}^2}^{s_0}ds  \left\{\frac{M_{\eta_c}^8}{4m_c^4} + \left(\frac{s}{2}-M_{\eta_c}^2\right)^2-\frac{M_{\eta_c}^4}{m_c^2} \left(\frac{s}{2}-M_{\eta_c}^2\right) \right\}\sqrt{1-\frac{4M_{\eta_c}^2}{s}} \,s\, \exp\left(-\frac{s}{T^2}\right)\nonumber\\
&&+\frac{f_{J/\psi}^4}{16\pi^2}\int_{4M_{J/\psi}^2}^{s_0}ds   M_{J/\psi}^4 \left\{2 +\frac{ \left(\frac{s}{2}-M_{J/\psi}^2\right)^2}{M_{J/\psi}^4} \right\} \sqrt{1-\frac{4M_{J/\psi}^2}{s}} \,s\, \exp\left(-\frac{s}{T^2}\right) \nonumber\\
&&=  \,\int_{16m_c^2}^{s_0} ds \int_{z_i}^{z_f}dz \int_{t_i}^{t_f}dt \int_{r_i}^{r_f}dr\, \rho_S(s,z,t,r)  \,s\, \exp\left(-\frac{s}{T^2}\right) \, ,
\end{eqnarray}

\begin{eqnarray}
&&\lambda^2_{X}\, \exp\left(-\frac{M^2_{X}}{T^2}\right) \nonumber\\
&&+\frac{f_{\eta_c}^4}{16\pi^2}\int_{4M_{\eta_c}^2}^{s_0}ds  \left\{\frac{M_{\eta_c}^8}{4m_c^4} + \left(\frac{s}{2}-M_{\eta_c}^2\right)^2-\frac{M_{\eta_c}^4}{m_c^2} \left(\frac{s}{2}-M_{\eta_c}^2\right) \right\}\sqrt{1-\frac{4M_{\eta_c}^2}{s}}\, \exp\left(-\frac{s}{T^2}\right)\nonumber\\
&&+\frac{f_{J/\psi}^4}{16\pi^2}\int_{4M_{J/\psi}^2}^{s_0}ds   M_{J/\psi}^4 \left\{2 +\frac{ \left(\frac{s}{2}-M_{J/\psi}^2\right)^2}{M_{J/\psi}^4} \right\} \sqrt{1-\frac{4M_{J/\psi}^2}{s}}\, \exp\left(-\frac{s}{T^2}\right) \nonumber\\
&&= \, \int_{16m_c^2}^{s_0} ds \int_{z_i}^{z_f}dz \int_{t_i}^{t_f}dt \int_{r_i}^{r_f}dr\, \rho_S(s,z,t,r)  \, \exp\left(-\frac{s}{T^2}\right) \, .
\end{eqnarray}

In Fig.7, we plot the mass $M_{X(cc\bar{c}\bar{c},0^{++})}$ comes  from Eqs.(29-30) with variation of the Borel parameter $T^2$, from the figure, we can see that the values of the $M_{X(cc\bar{c}\bar{c},0^{++})}$ are rather stable with variation of the Borel parameter at the energy scale $\mu\geq 1.8\,\rm{GeV}$. Now we choose the special value $T^2=4.4\,\rm{GeV}^2$, and plot the mass $M_{X(cc\bar{c}\bar{c},0^{++})}$  with variation of the energy scale $\mu$ in Fig.8. From the figure, we can see that the mass $M_{X(cc\bar{c}\bar{c},0^{++})}$ increases monotonously and quickly with increase of the energy scale $\mu$ at the region $\mu\leq 1.8\,\rm{GeV}$, and decreases monotonously and slowly at the region $\mu\geq 2.0\,\rm{GeV}$. In the range $\mu=(1.8-2.0)\,\rm{GeV}$, the predicted mass $M_{X(cc\bar{c}\bar{c},0^{++})}$ is rather stable, $M_{X(cc\bar{c}\bar{c},0^{++})}=5.76\,\rm{GeV}$, which is below the threshold $2M_{\eta_c}=5966.8\,\rm{MeV}$. On the other hand, the pole residue $\lambda_{X(cc\bar{c}\bar{c},0^{++})}$ increases monotonously and quickly with increase of the energy scale $\mu$ at the region $\mu\geq 1.6\,\rm{GeV}$, no stable QCD sum rules can be obtained, see Fig.9.  So the QCD sum rules cannot be saturated by the meson pairs $\eta_c\eta_c$, $J/\psi J/\psi$ plus a scalar tetraquark state $X(cc\bar{c}\bar{c},0^{++})$.

In this article, the
diquark-antidiquark type tetraquark state is taken as a special superposition of a series of  meson-meson pairs, and embodies  the net effects. The decays to its components (meson-meson pairs) are Okubo-Zweig-Iizuka super-allowed,  but the re-arrangements in the color-space are highly non-trivial. In other words, the lowest states $X(QQ\bar{Q}\bar{Q},0^{++}/2^{++})$ can saturate the QCD sum rules satisfactorily.

\begin{figure}
 \centering
 \includegraphics[totalheight=6cm,width=8cm]{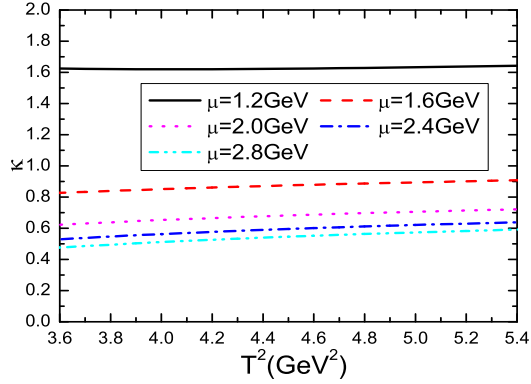}
        \caption{ The coefficient $\kappa$  with variation of the Borel parameter $T^2$.  }
\end{figure}

\begin{figure}
 \centering
 \includegraphics[totalheight=6cm,width=8cm]{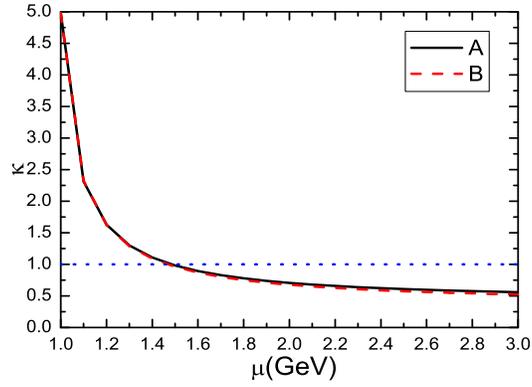}
        \caption{ The coefficient $\kappa$  with variation of the energy scale $\mu$, where $A$ and $B$ denote the values come from Eq.(27) and Eq.(28), respectively.  }
\end{figure}

\begin{figure}
 \centering
 \includegraphics[totalheight=6cm,width=8cm]{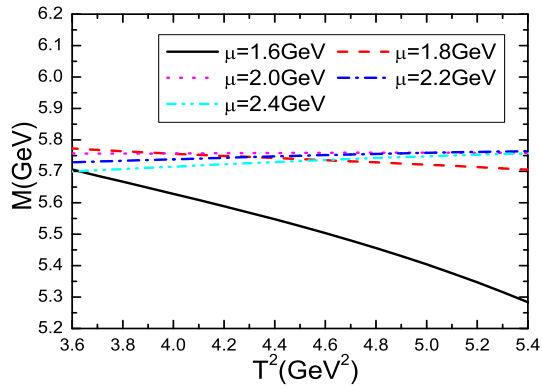}
        \caption{ The mass $M_{X(cc\bar{c}\bar{c},0^{++})}$ with variation of the Borel parameter $T^2$ from Eqs.(29-30).  }
\end{figure}

\begin{figure}
 \centering
 \includegraphics[totalheight=6cm,width=8cm]{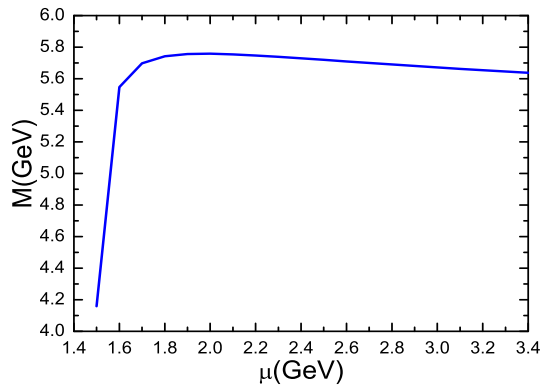}
        \caption{ The mass $M_{X(cc\bar{c}\bar{c},0^{++})}$   with variation of the energy scale $\mu$ from Eqs.(29-30).  }
\end{figure}

\begin{figure}
 \centering
 \includegraphics[totalheight=6cm,width=8cm]{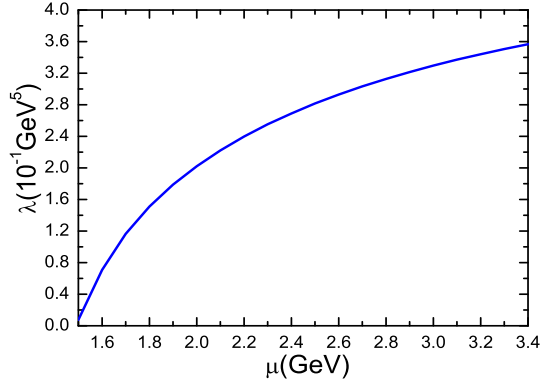}
        \caption{ The residue $\lambda_{X(cc\bar{c}\bar{c},0^{++})}$   with variation of the energy scale $\mu$ from Eqs.(29-30).  }
\end{figure}

\section{Conclusion}
In this article, we study the $J^{PC}=0^{++}$ and $2^{++}$ $QQ\bar{Q}\bar{Q}$ tetraquark states with the QCD sum rules by constructing  the diquark-antidiquark type   currents and calculating the contributions of the vacuum condensate up to dimension 4 in the operator product expansion. We obtain the predictions
$M_{X(cc\bar{c}\bar{c},0^{++})} =5.99\pm0.08\,\rm{GeV}$, $M_{X(cc\bar{c}\bar{c},2^{++})} =6.09\pm0.08\,\rm{GeV}$,
$M_{X(bb\bar{b}\bar{b},0^{++})} =18.84\pm0.09\,\rm{GeV}$, $M_{X(bb\bar{b}\bar{b},2^{++})}  =18.85\pm0.09\,\rm{GeV}$, which can be confronted to the experimental data in the future. Furthermore, we illustrate that the
diquark-antidiquark type tetraquark state can be  taken as a special superposition of a series of  meson-meson pairs and embodies  the net effects. We can search for the $J^{PC}=0^{++}$ and $2^{++}$ $QQ\bar{Q}\bar{Q}$ tetraquark states in the mass spectrum of the $\mu^+\mu^- \mu^+\mu^-$.

\section*{Appendix}
The explicit expression of the QCD spectral densities  $\rho_{S/T}(s)$,

\begin{eqnarray}
\rho_S(s,z,t,r)&=& \frac{3m_Q^4}{8\pi^6}\left( s-\overline{m}_Q^2\right)^2+\frac{t z m_Q^2}{8\pi^6}\left( s-\overline{m}_Q^2\right)^2\left( 5s-2\overline{m}_Q^2\right) \nonumber\\
&&+\frac{rtz(1-r-t-z)}{1-t-z} \frac{1}{32\pi^6}\left( s-\overline{m}_Q^2\right)^3\left( 3s-\overline{m}_Q^2\right) \nonumber\\
&&+\frac{rtz(1-r-t-z)}{1-z} \frac{1}{32\pi^6}\left( s-\overline{m}_Q^2\right)^3\left( 3s-\overline{m}_Q^2\right)\left[5-\frac{t}{1-t-z} \right] \nonumber\\
&&-\frac{rtz^2(1-r-t-z)}{1-z} \frac{3}{16\pi^6}\left( s-\overline{m}_Q^2\right)^4 \nonumber\\
&&+rtz(1-r-t-z) \frac{3s}{8\pi^6}\left( s-\overline{m}_Q^2\right)^2\left[ 2s-\overline{m}_Q^2-\frac{z}{1-z}\left( s-\overline{m}_Q^2\right)\right] \nonumber\\
&&+m_Q^2\langle \frac{\alpha_sGG}{\pi}\rangle \left\{-\frac{1}{r^3} \frac{m_Q^4}{6\pi^4}\delta\left( s-\overline{m}_Q^2\right) -\frac{1-r-t-z}{r^2} \frac{m_Q^2}{12\pi^4}\left[2+s\,\delta\left( s-\overline{m}_Q^2\right)\right]\right. \nonumber\\
&&-\frac{tz}{r^3} \frac{m_Q^2}{12\pi^4}\left[2+s\,\delta\left( s-\overline{m}_Q^2\right)\right]  -\frac{tz(1-r-t-z)}{r^2(1-t-z)} \frac{1}{12\pi^6}\left( 3s-2\overline{m}_Q^2\right) \nonumber\\
&&-\frac{tz(1-r-t-z)}{r^2(1-z)} \frac{1}{12\pi^4}\left( 3s-2\overline{m}_Q^2\right) \left[5-\frac{t}{1-t-z} \right]\nonumber\\
&&+\frac{tz^2(1-r-t-z)}{r^2(1-z)} \frac{1}{\pi^4}\left( s-\overline{m}_Q^2\right)  \nonumber\\
&&-\frac{tz(1-r-t-z)}{r^2} \frac{1}{2\pi^4}\left[s+\frac{s^2}{3}\delta\left( s-\overline{m}_Q^2\right)-\frac{z}{1-z}s\right]  \nonumber\\
&&\left.+\frac{1}{r^2} \frac{m_Q^2}{2\pi^4}   +\frac{tz}{r^2} \frac{1}{4\pi^4}\left( 3s-2\overline{m}_Q^2\right)-  \frac{1}{16\pi^4}\left( 3s-2\overline{m}_Q^2\right)  \right\} \nonumber
\end{eqnarray}
\begin{eqnarray}
&&+\langle \frac{\alpha_sGG}{\pi}\rangle \left\{\frac{1}{rz} \frac{m_Q^4}{6\pi^4} +\frac{t}{r} \frac{m_Q^2}{6\pi^4}\left( 3s-2\overline{m}_Q^2\right)\right. \nonumber\\
&&+\frac{t(1-r-t-z)}{(1-t-z)} \frac{1}{12\pi^4}\left( s-\overline{m}_Q^2\right) \left( 2s-\overline{m}_Q^2\right)  \nonumber\\
&&+\frac{t(1-r-t-z)}{(1-z)} \frac{1}{12\pi^4}\left( s-\overline{m}_Q^2\right) \left( 2s-\overline{m}_Q^2\right) \left[2-\frac{t}{1-t-z} \right] \nonumber\\
&&-\frac{tz(1-r-t-z)}{(1-z)} \frac{1}{4\pi^4}\left( s-\overline{m}_Q^2\right)^2\nonumber\\
&&\left.+t(1-r-t-z) \frac{1}{12\pi^4}s\left[ 4s-3\overline{m}_Q^2-\frac{z}{1-z}3\left( s-\overline{m}_Q^2\right)\right]\right\}\, ,
\end{eqnarray}

\begin{eqnarray}
\rho_T(s,z,t,r)&=& \frac{3m_Q^4}{16\pi^6}\left( s-\overline{m}_Q^2\right)^2+\frac{t z m_Q^2}{8\pi^6}\left( s-\overline{m}_Q^2\right)^2\left( 4s-\overline{m}_Q^2\right) \nonumber\\
&&+\frac{rtz(1-r-t-z)}{1-t-z} \frac{1}{320\pi^6}\left( s-\overline{m}_Q^2\right)^3\left( 17s-5\overline{m}_Q^2\right) \nonumber\\
&&+\frac{rtz(1-r-t-z)}{1-z} \frac{1}{320\pi^6}\left( s-\overline{m}_Q^2\right)^3\left[\left( 21s-5\overline{m}_Q^2\right)-\frac{t}{1-t-z}\left( 17s-5\overline{m}_Q^2\right) \right] \nonumber\\
&&-\frac{rtz^2(1-r-t-z)}{1-z} \frac{1}{32\pi^6}\left( s-\overline{m}_Q^2\right)^4 \nonumber\\
&&+rtz(1-r-t-z) \frac{s}{80\pi^6}\left( s-\overline{m}_Q^2\right)^2\left[ 28s-13\overline{m}_Q^2-\frac{z}{1-z}7\left( s-\overline{m}_Q^2\right)\right] \nonumber\\
&&+m_Q^2\langle \frac{\alpha_sGG}{\pi}\rangle \left\{-\frac{1}{r^3} \frac{m_Q^4}{12\pi^4}\delta\left( s-\overline{m}_Q^2\right) -\frac{1-r-t-z}{r^2} \frac{m_Q^2}{12\pi^4}\left[1+s\,\delta\left( s-\overline{m}_Q^2\right)\right]\right. \nonumber\\
&&-\frac{tz}{r^3} \frac{m_Q^2}{12\pi^4}\left[1+s\,\delta\left( s-\overline{m}_Q^2\right)\right]  -\frac{tz(1-r-t-z)}{r^2(1-t-z)} \frac{1}{12\pi^6}\left( 2s-\overline{m}_Q^2\right) \nonumber\\
&&-\frac{tz(1-r-t-z)}{r^2(1-z)} \frac{1}{12\pi^4}\left( 2s-\overline{m}_Q^2\right) \left[1-\frac{t}{1-t-z} \right]\nonumber\\
&&+\frac{tz^2(1-r-t-z)}{r^2(1-z)} \frac{1}{6\pi^4}\left( s-\overline{m}_Q^2\right)  \nonumber\\
&&-\frac{tz(1-r-t-z)}{r^2} \frac{1}{6\pi^4}\left[s+\frac{s^2}{2}\delta\left( s-\overline{m}_Q^2\right)-\frac{z}{1-z}s\right]  \nonumber\\
&&\left.+\frac{1}{r^2} \frac{m_Q^2}{4\pi^4}   +\frac{tz}{r^2} \frac{1}{4\pi^4}\left( 2s-\overline{m}_Q^2\right)  \right\} \nonumber
\end{eqnarray}
\begin{eqnarray}
&&+\langle \frac{\alpha_sGG}{\pi}\rangle \left\{-  \frac{m_Q^2}{48\pi^4}\left( 4s-3\overline{m}_Q^2\right)\right. \nonumber\\
&&-\frac{r(1-r-t-z)}{1-t-z}\frac{1}{32\pi^4} \left( s-\overline{m}_Q^2\right)\left( 3s-\overline{m}_Q^2\right)\nonumber\\
&&-\frac{r(1-r-t-z)}{1-z}\frac{1}{480\pi^4} \left( s-\overline{m}_Q^2\right)\left[\left( 17s-5\overline{m}_Q^2\right)-\frac{t}{1-t-z}15\left( 3s-\overline{m}_Q^2\right)\right]\nonumber\\
&&+\frac{rz(1-r-t-z)}{1-z}\frac{1}{24\pi^4} \left( s-\overline{m}_Q^2\right)^2 \nonumber\\
&&-r(1-r-t-z)\frac{1}{240\pi^4} s\left[\left( 14s-9\overline{m}_Q^2\right)-\frac{z}{1-z}21\left( s-\overline{m}_Q^2\right)\right]\nonumber\\
&&-\frac{1}{rz} \frac{m_Q^4}{36\pi^4} -\frac{t}{r} \frac{m_Q^2}{18\pi^4}\left( 2s-\overline{m}_Q^2\right)\nonumber\\
&&-\frac{t(1-r-t-z)}{(1-t-z)} \frac{1}{72\pi^4}\left( s-\overline{m}_Q^2\right) \left( 4s-\overline{m}_Q^2\right)  \nonumber\\
&&-\frac{t(1-r-t-z)}{(1-z)} \frac{1}{72\pi^4}\left( s-\overline{m}_Q^2\right)  \left[2\left( 2s-\overline{m}_Q^2\right)-\frac{t}{1-t-z}\left( 4s-\overline{m}_Q^2\right) \right] \nonumber\\
&&+\frac{tz(1-r-t-z)}{(1-z)} \frac{1}{24\pi^4}\left( s-\overline{m}_Q^2\right)^2\nonumber\\
&&\left.-t(1-r-t-z) \frac{1}{72\pi^4}s\left[ 7s-5\overline{m}_Q^2-\frac{z}{1-z}5\left( s-\overline{m}_Q^2\right)\right]\right\}\, ,
\end{eqnarray}
where
\begin{eqnarray}
\overline{m}_Q^2&=&\frac{m_Q^2}{r}+\frac{m_Q^2}{t}+\frac{m_Q^2}{z}+\frac{m_Q^2}{1-r-t-z}\, .
\end{eqnarray}

\section*{Acknowledgements}
This  work is supported by National Natural Science Foundation,
Grant Numbers 11375063,  and Natural Science Foundation of Hebei province, Grant Number A2014502017.


\begin{thebibliography}{99}


\bibitem{Zhu-PRT} H. X. Chen, W. Chen, X. Liu and S. L. Zhu, Phys. Rept. {\bf 639} (2016) 1;
A. Esposito, A. Pilloni and A. D. Polosa, Phys. Rept. {\bf 668} (2016) 1.

\bibitem{Silvestre-1986}   S. Zouzou, B. Silvestre-Brac, C. Gignoux and J. M. Richard, Z. Phys. {\bf C30} (1986) 457;
 C. Semay and B. Silvestre-Brac, Z. Phys. {\bf C61} (1994) 271.


 \bibitem{Lloyd-2004} R. J. Lloyd and J. P. Vary, Phys. Rev. {\bf D70} (2004) 014009.


 \bibitem{Barnea-2006} N. Barnea, J. Vijande and A. Valcarce, Phys. Rev. {\bf D73} (2006) 054004.

\bibitem{Bai-2016}  Y. Bai, S. Lu and J. Osborne, arXiv:1612.00012.


\bibitem{Heupel-2012}  W. Heupel, G. Eichmann and C. S. Fischer, Phys. Lett. {\bf B718} (2012) 545.


\bibitem{Berezhnoy-2012} A. V. Berezhnoy, A. V. Luchinsky and A. A. Novoselov, Phys. Rev. {\bf D86} (2012) 034004.

\bibitem{Rosner-2016} M. Karliner, J. L. Rosner and S. Nussinov,  Phys. Rev. {\bf D95} (2017)  034011.

\bibitem{Wu-2016} J. Wu, Y. R. Liu, K. Chen, X. Liu and S. L. Zhu, arXiv:1605.01134.


\bibitem{Chen-2016}  W. Chen, H. X. Chen, X. Liu, T. G. Steele and S. L. Zhu, arXiv:1605.01647.


\bibitem{QCDSR-Tetraquark-Molecule} R. D. Matheus, S. Narison, M. Nielsen and J. M. Richard, Phys. Rev. {\bf D75} (2007) 014005;
R. M. Albuquerque and M. Nielsen, Nucl. Phys. {\bf A815} (2009) 53;
Z. G. Wang, Eur. Phys. J. {\bf C63} (2009) 115;
J. R. Zhang and M. Q. Huang, Commun. Theor. Phys. {\bf 54} (2010) 1075;
Z. G. Wang, Eur. Phys. J. {\bf C67} (2010) 411;
W. Chen and S. L. Zhu, Phys. Rev. {\bf D83} (2011) 034010;
J. R. Zhang, M. Zhong and M. Q. Huang, Phys. Lett. {\bf B704} (2011) 312;
Z. G. Wang and T. Huang, Phys. Rev. {\bf D89} (2014)  054019;
Z. G. Wang, Eur. Phys. J. {\bf C74} (2014)  2874;
Z. G. Wang and T. Huang, Nucl. Phys. {\bf A930} (2014) 63;
S. S. Agaev, K. Azizi and H. Sundu, Phys. Rev. {\bf D93} (2016) 114007;
Z. G. Wang, Eur. Phys. J. {\bf C76} (2016)  279.


\bibitem{Exp-psipsi} R. Aaij et al, Phys. Lett. {\bf B707} (2012) 52;
V. Khachatryan et al, JHEP {\bf 1409} (2014)  094;
M. Aaboud et al, Eur. Phys. J. {\bf C77} (2017)  76.


\bibitem{Exp-UpsilonUpsilon} V. Khachatryan et al, JHEP {\bf 1705} (2017) 013.


\bibitem{One-gluon} A. De Rujula, H. Georgi and S. L. Glashow, Phys. Rev.  {\bf D12} (1975) 147;
 T. DeGrand, R. L. Jaffe, K. Johnson and J. E. Kiskis, Phys.  Rev.  {\bf D12} (1975) 2060.

\bibitem{WangDiquark} Z. G. Wang, Eur. Phys. J. {\bf C71} (2011) 1524;
  R. T. Kleiv, T. G. Steele and A. Zhang, Phys. Rev. {\bf D87} (2013) 125018.

\bibitem{WangLDiquark}  Z. G. Wang, Commun. Theor. Phys. {\bf 59} (2013) 451.


\bibitem{SVZ79}  M. A. Shifman, A. I. Vainshtein and V. I. Zakharov, Nucl. Phys. {\bf B147} (1979) 385;
Nucl. Phys. {\bf B147} (1979) 448.

\bibitem{Reinders85} L. J. Reinders, H. Rubinstein and S. Yazaki, Phys. Rept. {\bf 127} (1985) 1.


\bibitem{Wang4430} Z. G. Wang,  Commun. Theor. Phys. {\bf 63} (2015) 325.

\bibitem{Wang4500} Z. G. Wang, Eur. Phys. J. {\bf C77} (2017) 78.

\bibitem{ColangeloReview} P. Colangelo and A. Khodjamirian, hep-ph/0010175.

\bibitem{PDG}  C. Patrignani et al, Chin. Phys. {\bf C40} (2016)  100001.


\bibitem{Wang-BC} Z. G. Wang, Eur. Phys. J. {\bf A49} (2013) 131.


\bibitem{WangD3K} Z. G. Wang, Nucl. Phys. {\bf A957} (2017) 85.


\bibitem{RevWise} A. V. Manohar and M. B. Wise, Camb. Monogr. Part. Phys. Nucl. Phys. Cosmol. {\bf 10} (2000) 1.


\bibitem{SELEX2002}  M. Mattson et al,  Phys. Rev. Lett. {\bf 89} (2002) 112001.

\bibitem{SELEX2004} A. Ocherashvili  et al,  Phys. Lett.   {\bf B628} (2005) 18.


\bibitem{Wang-Xcc} Z. G. Wang, Eur. Phys. J. {\bf A45} (2010) 267.


\bibitem{feta-fpsi} D. Becirevic, G. Duplancic, B. Klajn, B. Melic and F. Sanfilippo, Nucl. Phys. {\bf B883} (2014) 306.



\end{thebibliography}
\end{document}